%% file: main.tex
\documentclass[10pt,letterpaper]{article}
\usepackage[top=0.85in,left=2.75in,footskip=0.75in]{geometry}

\usepackage{amsmath,amssymb}

\usepackage{changepage}
\usepackage{comment}

\usepackage{listings}

\usepackage{textcomp,marvosym}

\usepackage{cite}

\usepackage{nameref,hyperref}

\usepackage[nopatch=eqnum]{microtype}
\DisableLigatures[f]{encoding = *, family = * }

\usepackage[table, dvipsnames]{xcolor}

\usepackage{array}

\newcolumntype{+}{!{\vrule width 2pt}}

\newlength\savedwidth

\raggedright
\usepackage[aboveskip=1pt,labelfont=bf,labelsep=period,justification=raggedright,singlelinecheck=off]{caption}

\makeatletter
\renewcommand{\@biblabel}[1]{\quad#1.}
\makeatother

\usepackage{lastpage,fancyhdr,graphicx}
\usepackage{epstopdf}
\fancyheadoffset[L]{2.25in}
\fancyfootoffset[L]{2.25in}
\usepackage[utf8]{inputenc}
\usepackage[T1]{fontenc}
\usepackage{comment}
\usepackage{xcolor}
\usepackage[binary-units]{siunitx}
\usepackage[]{graphicx}
\usepackage{xurl}
\usepackage{authblk}
\usepackage{subcaption}
\usepackage[framemethod=TikZ]{mdframed}

\input{code-style}

\definecolor{my_bg}{HTML}{F0F0F0}
\definecolor{my_title_bg}{HTML}{383838}
\newcounter{Frame}

\begin{document}
\input{macros}

% \title{Scaling Science: Approaches and Experiences Applying Event-driven Methods to Distributed Research Computing}

\begin{flushleft}{\Large\textbf\newline{AERO: An autonomous platform for continuous research} 
}
\newline

%\title{}

\input{authors}

\end{flushleft}
%\maketitle

\input{abstract}

\input{sections/intro-alt}

\input{sections/methods}
\input{sections/evaluation}
\input{sections/discussion}
%\input{sections/background}

% \subsection{Background}
% \subsubsection{Globus Auth}
% \subsubsection{Globus Compute}
% \subsubsection{Globus Connect Server}
% \subsubsection{Globus Flows}
% \subsubsection{Globus Search}
% \subsubsection{Globus Timers}

% \input{sections/implementation}

% \input{sections/evaluation}

% \input{sections/supplementary}

% \input{sections/related_work}

%\input{sections/conclusion}

\section{Code Availability}
\name{} server implementation code is available at \url{https://github.com/NSF-RESUME/aero}. Dockerfiles are provided for convenience, with a Docker setup script provided in \url{https://github.com/NSF-RESUME/aero/scripts/prepare_start.sh}. The client code is available at \url{https://github.com/NSF-RESUME/aero-client}. R(t) calculation code on hospitalization records is available at \url{https://github.com/NSF-RESUME/reff-chicago/} and R(t) calculation on wastewater data is available at \url{https://github.com/kyleconroy3/Chicago-WW-Reff.git}

\section{Data Availability}
Data ingested for the clinical surveillance data flow is available at \url{https://data.cityofchicago.org/Health-Human-Services/COVID-19-Daily-Cases-Deaths-and-Hospitalizations-H/naz8-j4nc/about_data}. Data ingested for the analysis of pathogen concentrations in wastewater is available at \url{https://iwss.uillinois.edu/wastewater-treatment-plant/169/?target=SARS-CoV-2}. All \name{} monitored data generated by the flows can be found in the publicly available Globus Collection \url{https://app.globus.org/file-manager?destination_id=52f7f6bc-444f-439a-ad48-a4569d10c3d1&destination_path=%2F}.

\section{Author Contributions}
VHS, JO, KCh., AS, NC and YB contributed to conceptualization and project design. VHS, SS and JGP contributed to the development of the platform front and backend. AS, NC, KCo. contributed to the development of the $R(t)$ use case code. VHS and NH contributed to code documentation. VHS, JO, KCh., AS, NC, JGP, IF, KCo., MG, NH and DDSG contributed to the writing and editing of the manuscript.

\section{Competing Interests}
The authors declare no competing interests.

\input{sections/acknowledgements}



\bibliography{bib}

\end{document}

%% file: code-style.tex
\makeatletter
\apptocmd\@makecaption{\par}{}{%
  \errmessage{\noexpand\@makecaption could not be patched}%
}
\makeatother

\usepackage{inconsolata}

\definecolor{code_bg}{RGB}{239, 241, 245}
\definecolor{code_comment}{RGB}{79, 79, 105}
\definecolor{code_string}{RGB}{64, 160, 43}
\definecolor{code_keyword}{RGB}{210, 15, 57}
\definecolor{code_number}{RGB}{254, 100, 11}

%% file: macros.tex
% comments
\newcommand\valerie[1]{{\color{blue}{Valerie: #1}}}
\newcommand\kyle[1]{{\color{red}{Kyle: #1}}}
\newcommand\JO[1]{{\color{red}{Jonathan: #1}}}
\newcommand\abby[1]{{\color{purple}{Abby: #1}}}
\newcommand\ian[1]{{\color{orange}{Ian: #1}}}
% names and acronyms
\newcommand{\name}[0]{\texttt{AERO}}
\newcommand{\dsaas}{AERO}
\newcommand{\globus}{Globus}
\newcommand{\gc}{Globus Compute}
\newcommand{\gcs}{GCS}
\newcommand{\osprey}{OSPREY}

%%%%%%%%%%%%%%%%%%%%%%%%%%%%%%%%%%%%%%%%%%%%%%%%%%%%%%%%%%%%%%%%%%%%%%%%%%%%%%
%%%%%%%%%%%%%%%%%%%%%%%%%%%%%%%%%%%%%%%%%%%%%%%%%%%%%%%%%%%%%%%%%%%%%%%%%%%%%%

%% file: authors.tex
Val\'erie Hayot-Sasson\textsuperscript{1,2},
Abby Stevens\textsuperscript{2},
Nicholson Collier\textsuperscript{2},
Sudershan Sridhar\textsuperscript{3},
Kyle Conroy\textsuperscript{1},
J. Gregory Pauloski\textsuperscript{1,2},
Yadu Babuji\textsuperscript{1},
Maxime Gonthier\textsuperscript{1,2},
Nathaniel Hudson\textsuperscript{1,2},
Dante D. Sanchez-Gallegos\textsuperscript{4},
Ian Foster\textsuperscript{1,2},
Jonathan Ozik\textsuperscript{1,2},
Kyle Chard\textsuperscript{1,2}
\\
\bigskip
\textbf{1} University of Chicago, Chicago, Illinois, USA.
\\
\textbf{2} Argonne National Laboratory, Lemont, Illinois, USA.
\\
\textbf{3} Globus, Chicago, Illinois, USA.
\\
\textbf{4} University Carlos III of Madrid, Leganes, Madrid, Spain.
\\
\bigskip

% \author[1,2]{Val\'erie Hayot-Sasson}
% \author[2]{Abby Stevens}
% \author[2]{Nicholson Collier}
% \author[3]{Sudershan Sridhar}
% \author[1]{Kyle Conroy}
% \author[1,2]{J. Gregory Pauloski}
% \author[1]{Yadu Babuji}
% \author[1,2]{Maxime Gonthier}
% \author[1,2]{Nathaniel Hudson}
% \author[4]{Dante D. Sanchez-Gallegos}
% \author[1,2]{Ian Foster}
% \author[1,2]{Jonathan Ozik}
% \author[1,2]{Kyle Chard}

% \affil[1]{University of Chicago, Chicago, Illinois, USA.}
% \affil[2]{Argonne National Laboratory, Lemont, Illinois, USA.}
% \affil[3]{Globus, Chicago, Illinois, USA.}
% \affil[4]{University Carlos III of Madrid, Leganes, Madrid, Spain.}

%% file: abstract.tex
\section*{Abstract}
% \valerie{170 words max}
% \ian{I propose here a rewrite of the abstract. Original is commented out in the text.}
The COVID-19 pandemic highlighted the need for new data infrastructure, as epidemiologists and public health workers raced to harness rapidly evolving data, analytics, and infrastructure in support of cross-sector investigations.
To meet this need, we developed \name{}, an automated research and data sharing platform for continuous, distributed, and multi-disciplinary collaboration.
In this paper, we describe the \name{} design and how it supports the automatic ingestion, validation, and transformation of monitored data into a form suitable for analysis; the automated execution of analyses on this data; and the sharing of data among different entities. 
We also describe how our \name{} implementation leverages capabilities provided by the Globus platform and GitHub for automation, distributed execution, data sharing, and authentication.
% We present results obtained from running three applications (a synthetic application and two COVID-19 monitoring applications), which is publicly available for testing.
We present results obtained with an instance of \name{} running two public health surveillance applications and demonstrate benchmarking results with a synthetic application, all of which are publicly available for testing.
%Motivated by gaps in data, analytics, infrastructure and cross sector collaborations identified during the COVID-19 pandemic, we sought to create a hub to address the gaps occurring in such rapidly evolving scenarios. Such a solution would require mechanisms to automatically ingest, validate and transform the data to ensure it is viable for analysis. Additionally, mechanisms to support automated analyses on monitored data and mechanisms to share monitored data between different entities are also required. In this paper, we present \name{}, an automated research and data sharing hub, that integrates various Globus services to provide automation and distributed execution, data sharing and authentication. \name{} has been developed as a prototype running three applications (a synthetic application and two COVID-19 monitoring applications) and is publicly available for testing.
% Long-running experiments are typically composed of a series of workflows glued together by
% events. Examples of such events include data being made available and results meeting a certain threshold. As new data is made available, the same subset of events will trigger and launch the same workflows, given reason for automation. To address the needs for automation in scientific workflows, we developed \dsaas, a platform for both data sharing and 

%% file: sections/intro-alt.tex
\section*{Introduction}

The research process is composed of many sequential and iterative steps.
For example, in observational science, researchers analyze rapidly evolving data by applying sophisticated data processing workflows, comparing observed data to model outputs, and exploring the potential impact of various interventions.   
% they must first wait for this data to be produced and shared, before launching analyses.
The results of these analyses may lead to further questions, resulting in acquisition of new data or execution of new analyses. These processes typically include a ``human-in-the-loop,'' for example waiting for sufficient data to be collected before performing analysis, running and monitoring analyses, interpreting results, and determining the next steps to follow. 
%analysing the results and potentially only then implementing and executing further analysis on this data. 
Further automating these aspects of the research process can both reduce overheads on researchers---allowing  their time to be better spent on analysis and exploration---and also reduce time to discovery.

% \kyle{We might need to move this para as it sounds like more intro}
% Much of today's scientific research is oriented around data. Data are used to pose hypotheses, conduct experiments, and prove or disprove hypotheses leading to new knowledge. Data are obtained from many sources: observational data from sensors, published data from publications or data repositories, simulated data from models, and experiment data from instruments. Making use of this data requires sophisticated processes to ingest, validate, and transform data before being combined with other data for analysis. The scientific process is cyclical in that results from analyses inform new questions, necessitating acquisition of new data from diverse sources or producing new data that challenges previous assumptions.
% \valerie{something here from jonathan about the covid 19 pandemic to help with flow}
The COVID-19 pandemic highlighted the critical need for increased automation as epidemiologists faced a deluge of potentially useful data, new epidemiological models, and unparalleled computing resources. 
% manually monitored rapidly updated data, adapted analyses to new data sources, and experimented with different model configurations. 
%Analysis of important data was reliant on humans to obtain, validate, integrate, and process data. 
Unfortunately, in spite of these significant resources, humans quickly became the bottleneck in this process. 
%in the research process.  
Consider, for example, the following vignette---a common scenario for many researchers around the world: 
% For instance, a standard set of workflows in disease monitoring may be triggered by the following events:
1) data published by a hospital or health authority are monitored for updates; 2) data are reviewed for quality and validated according to various assumptions; 3) data are used to calibrate models and investigate different scenarios; 4) model results are used by decision makers to inform their policies. 
%, used to train new  models, or replicated on persistent storage.
The binding elements between these steps are the researchers themselves, who are informed that new data or results have been obtained and must make decisions as to which step to execute next. The sequence of steps that are to be taken may be known in advance, providing an opportunity for significant aspects of the process to be automated, requiring only human intervention when unexpected events occur. % are obtained.

% Manually managing such workflows is cumbersome and prone to failure. As a result, 
Researchers have long sought to automate specific parts of the research process. For example, workflow management systems~\cite{zhao2005notation, wilde2009parallel, babuji2019parsl, deelman2015pegasus, di2017nextflow, ozik_desktop_2016, collier_distributed_2024} are widely used to define and manage the execution of a sequence of computational tasks~\cite{babuji2019parsl} and other actions (e.g., data movement, publication) across systems~\cite{vescovi22linking, chard23automation}. Data analysis workflows rely on a standardized set of preprocessing steps to be performed prior to analysis. With large-scale analyses, 
%(e.g., public health surveillance, high energy physics, genomics)
these steps need to be repeated as new data are added. For experiments that rely on data acquired from many sites, users need not only to repeat preprocessing steps as new data are received, but also to adapt preprocessing for different data sources such that data are in the correct formats for combined analyses. Due to the standardized nature of data preprocessing, automation has the potential to increase the efficiency and reliability of data ingestion and integration. However, computational analyses that rely on data can vary widely. In the public health use case that we highlight, there are many modeling methods, spatiotemporal scales, analysis goals, and computational requirements. Developing automation capabilities that can support the different and internally diverse research communities requires an eye towards extensibility and robustness.

Automation has previously been incorporated into scientific research workflows, particularly for public health modeling. 
%however, the amount of automation is limited. 
For example, Hubverse~\cite{hubverse} and FluSight~\cite{flusight} %are two projects aimed at 
focus on monitoring disease progression over time. These ``active'' repositories %Projects such as these 
are the culmination of many diverse \textit{workflows} that are run as a result of various \textit{events}. Nevertheless, these projects focus primarily on data sharing, providing automation for shared data uploaded via GitHub Actions and are tied to a specific domain, limiting their applicability to other research areas which can also benefit from automation.

We present an open-source hybrid and asynchronous data research automation platform called Automated Event-based Research Orchestration (\name{}). 
\name{} is a cloud-hosted event-based %data and analysis 
platform that allows researchers to automate arbitrary research processes via a novel trigger-action paradigm. 
%automates execution of workflows %and actions 
%as the result of events. 
\name{} builds on Globus~\cite{chard14globus} and other cloud services for security, data management, workflow execution, and data access. % results available. 
\name{} is implemented as a distributed platform, storing metadata centrally and integrating distributed user-owned and -managed resources (e.g., laptops, clouds, clusters) for data storage and workflow execution. 

We first describe our motivating use case, highlighting the challenges faced by researchers responding
to the COVID-19 pandemic. We then describe the AERO platform, 
presenting critical requirements derived from our experiences, and outlining the design decisions that informed
the creation of AERO. We then present our prototype implementation that is publicly deployed for use in public health decision support. We evaluate the platform using a synthetic 
application and two real-world modeling applications. Our results demonstrate the performance and utility of AERO. We conclude by discussing our experiences and outlining our vision for future work in this area.

\refstepcounter{Frame}
\begin{mdframed}
[%
    frametitle={%
        \textcolor{white}{Frame \theFrame. Motivating use case: Lessons from COVID-19 modeling}
    },
    skipabove=\baselineskip plus 2pt minus 1pt,
    skipbelow=\baselineskip plus 2pt minus 1pt,
    linewidth=0.5pt,
    frametitlerule=true,
    frametitlebackgroundcolor=my_title_bg,
    backgroundcolor=my_bg,
    roundcorner=10pt,
    middlelinecolor=my_title_bg,
    middlelinewidth=1pt,
]
% \begin{Frame}[Motivating use case: Lessons from COVID-19 modeling]
\label{frame:motivation}

% \kyle{First pass. Jonathan - would be good to have you take a look and see if you agree/add some examples if you can. }
% \valerie{put a box around this. was thinking we could have all the use case-related stuff in boxes to highlight them, but maybe a bad idea}
% \JO{I added and edited the text below, see Overleaf comments.}

The COVID-19 pandemic revealed fundamental gaps in data, analytics, infrastructure, and cross-sector
collaborations needed to effectively monitor and respond to rapidly evolving health threats.

\vspace{1ex}

New data sources became available throughout the pandemic that were found to provide new insight into its evolution and the efficacy of policies. Data sources grew to include cases, hospitalizations, hospital statistics (e.g., beds and ventilators), wastewater, mobility, survey, vaccination, therapeutics, sensor networks, and many others. These data were released by many different authorities (e.g., every hospital, county, city, state) using different web-based interfaces, protocols, authentication/authorization models, and data formats. This heterogeneity presented a significant problem to modelers who aimed to integrate and utilize new sources as quickly as possible. Without a common platform, they relied on ad hoc and manual methods to determine when data were updated and to download those data for local use. 

\vspace{1ex}

Data were heterogeneous, changing, incomplete, flawed, and delayed. As such, epidemiologists developed pipelines for ingesting, validating, and transforming the data. These pipelines were responsible for ingesting the heterogeneous formats, converting them to a common representation, and importantly, performing quality control. They discovered that column names and vocabularies were changed over time, that classifications were modified as the pandemic proceeded (e.g., what was considered a positive case), and in some cases data that was assumed to be monotonically increasing (e.g., total cases, hospitalizations, and deaths) decreased due to review of historical data or different classifications. These types of changes presented significant challenges to users of the data and required close scrutiny of data used for modeling and analysis.

\vspace{1ex}

Researchers used these data to inform, constrain, and predict epidemic trends through computational analyses utilizing a variety of modeling methods (statistical~\cite{leisman_modeling_2024}, compartmental~\cite{runge_modeling_2022}, meta-population~\cite{lima_value_2024}, agent-based~\cite{ozik_population_2021}), focused on different geographic extents (city~\cite{hotton_impact_2022,chang_mobility_2021}, county~\cite{shea_multiple_2023}, state~\cite{bollyky_assessing_2023}, national~\cite{watson_global_2022}), temporal scales (short-term forecast~\cite{ray_ensemble_2020}, medium/long-term planning~\cite{borchering_modeling_2021}), and outcomes (cases, hospital capacity, non-health impacts~\cite{lima_reopening_2021}).  They developed their own, primarily manual, methods for building and validating their analyses, updating them as data were updated. Given the significant individual efforts involved, capabilities that would have facilitated these analyses to mutually inform and be combined could have yielded improved support of decision making during different stages of the unfolding public health emergency. The computational requirements of some of the epidemic analyses were modest, e.g., computing of the effective reproduction number ($R(t)$), a common epidemiological metric used to measure changes in disease transmission~\cite{gostic_practical_2020}, from hospitalizations and emergency room visits~\cite{richardson_tracking_2022}. However, extending these to many geographic regions, incorporating uncertainties in the underlying assumptions, or utilizing noisier, passive surveillance signals such as those from wastewater~\cite{goldstein_semiparametric_2024}, could rapidly increase the computational requirements. Incorporating mechanistic models, where large numbers of simulations are used to understand epidemic trends, further exacerbates the computing needs, often requiring high-performance computing resources~\cite{ozik_population_2021}.

\vspace{1ex}

Despite the unprecedented resources allocated to COVID-19 research, researchers primarily worked independently to analyze data and provide epidemiological model output. As a result, researchers developed similar infrastructures for managing large data and compute tasks when developing, verifying, and applying their models. 
While communication between research groups and public health stakeholders was at unprecedented levels during the pandemic, there was surprisingly little sharing of methods, data, and results in spite of the obvious benefits of integrating analysis methods to provide a broader view of the pandemic and exploiting new data sources as they became available. Researchers shared data sources via out-of-band methods, websites, and GitHub repositories; similarly, analyses and underlying epidemiological models were often shared via source code. In the later stages of the pandemic, researchers developed services for publishing and comparing specific model types. This enabled comparison between model outputs and evaluation of the best performing methods.
% \end{mybox}
% \end{Frame}
\end{mdframed}

%The one-off approaches risk sacrificing robustness, reproducibility,
%security, scalability, and efficiency. There is a fundamental need to drive innovations in infrastructure for  predictive intelligence for pandemic prevention (PIPP) data and analytics, to lower the barriers to andautomate surveillance, epidemiological analyses, and rapid response on cloud and HPC resources. To be transformative, infrastructure must provide accessible capabilities developed through best practices and informed by data and model safety and ethics principles22.

%% file: sections/methods.tex
\section*{Methods}
We designed Automated Event-based Research Orchestrator, or \textit{\name{}}, to automate a range of common activities found in many scientific domains. 
%address the automation needs of the various long-running analyses found in many scientific domains. %\name{} was designed to satisfy these requirements and enable automated scientific computing. 
Here we describe important requirements for automating scientific research. We then describe how \name{} addresses these requirements, present \name{}'s architecture,  and explain how \name{} can be used for analysis.

\subsection*{Automated science requirements}

% \begin{table}[]
%     \centering
%     \caption{Automated scientific research requirements}
%     \label{tab:requirements}
%     \begin{tabular}{p{5.7cm}|p{6cm}}
%          Requirement & Solution  \\
%          \hline
%          Data updates occur at varying intervals & Triggers to fetch data on updates \\
%          Wide range of data sources & Ability to fetch data from different locations using various protocols and authorization requirements \\
%          Data contents evolve and change over time & Ability to run automated workflows that validate and transform data \\
%     \end{tabular}
% \end{table}
%\valerie{Take two on explaining the requirements}
Our motivating use case (\autoref{frame:motivation}) elucidates critical requirements necessary for the automation of scientific applications, spanning hardware, data sharing and provenance. %These requirements can be separated into broadly two categories, 1) automation and 2) data sharing and provenance.
More precisely, the requirements necessary for automated execution include:
% \begin{itemize}
%     \item Execution of arbitrary functions on arbitrary hardware
%     \item Flexible trigger-event rules describing various conditions
%     \item Error handling and retries
% \end{itemize}

\textbf{Arbitrary functions and hardware:} Scientific applications can be described using programming languages or domain-specific languages (DSL). To avoid limiting the platform to a subset of scientific applications or enforcing the rewrite of scientific applications, \name{} must be language-agnostic.
%Furthermore, scientific applications may require access to specialized hardware (e.g., access to GPUs for machine learning applications). While many solutions opt to adhere to specific sites or cloud providers, our platform needs to support all infrastructure a.

In addition, scientific applications may require access to specialized infrastructure or must be able to make use of available resources at a given time. Providing a subset of compute infrastructure through cloud resources within \name{} may not satisfy application demands. Therefore, \name{} must provide users with the ability to use their existing infrastructure for automation.

\textbf{Trigger-event rules: } \name{} should support the expression of many different rules to trigger automation of scientific processes. Broadly speaking, scientific applications may need to be executed when 1) application data are modified or 2) periodically when application data is fetched from external sources. Input data may be updated when new versions are published or when computational models have been updated. The flows that depend on these inputs may choose to wait for some or all of the inputs to be updated. If inputs need to be gathered from external repositories, periodic triggers are used to gather regular updates from external sources.

\textbf{Error handling and retries:} Scientific flow execution may fail for a variety of reasons, including both transient and terminal errors. \name{} must be able to detect and recover from these errors. For example, terminal errors may arise from updates in metadata, data or external source locations and must be relayed back to the user such that necessary actions can be taken. For transient errors, it may be more effective to first attempt to retry the flow, rather than immediately informing the user of the error.

\textbf{Data management and sharing: }
% \hfill\break
Data processed and shared by scientific applications have their own set of requirements. %These requirements include:
 % \begin{itemize}
 %     \item Data can only be accessed by authorized users
 %     \item Data should adhere to the 
 % \end{itemize}
For example, not all scientific data can be widely shared. Private scientific data can include proprietary data with restricted access, including Personally Identifiable Information (PII), which can contain non-anonymized features. As a result, \name{} must ensure that access be restricted to some data. 
%scientific data sharing platforms to restrict access to certain data to only authorized users. 
Moreover, to facilitate collaboration and further automation of data analysis, \name{} must also make it possible for shared data to adhere to the FAIR~\cite{wilkinson2016fair} (findable, accessible, interoperable and reproducible) principles. These principles offer a set of guidelines to ensure that digital assets are machine actionable with limited need for human interaction.

\subsection*{Automated Event-based Research Orchestrator}
% \valerie{maybe unclear what events are?}

\begin{figure}[h]
    \centering
    \includegraphics[width=1\linewidth]{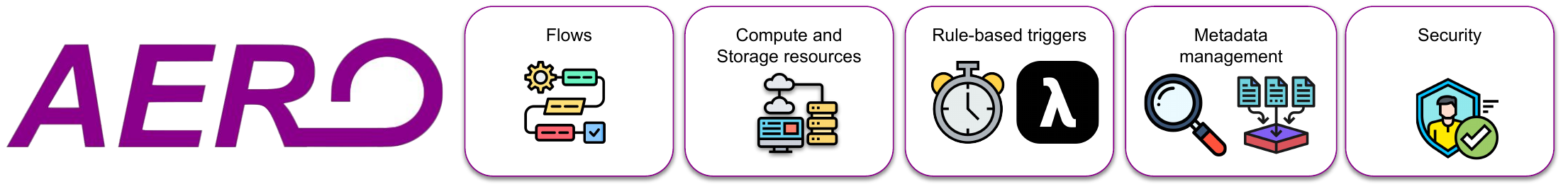}
    \caption{\name{} enables automated workflow execution by incorporating five core components. Users interact with the \textit{metadata} server to register \textit{flows} using \textit{rule-based triggers}, which drive the automation. When \textit{rules} are satisfied, they trigger user-defined \textit{actions} that execute within predefined ingestion or analysis \textit{flows} on user-provided \textit{compute and storage resources}. As \name{} leverages user-provided data and infrastructure, a \textit{security} model is used to ensure that access to unauthorized resources is forbidden.}\label{fig:dsaas}
\end{figure}
% \valerie{distributed compute, user-defined ingestion and analysis instead of flows, rich metadata and provenance + version (FAIR)}

%To meet the various needs of automated scientific research, \name{} was developed. 
\name{} is a collaborative automation hub that enables users to define automated \textit{flows} that are executed based on \textit{events} (e.g., data updates). The data resulting from the execution of automated flows can be monitored by \name{}, thus enabling the chaining of various flows. \name{} also provides data sharing and discovery capabilities, allowing users to share and build upon existing flows within and across teams.

As shown in Figure~\ref{fig:dsaas}, \name{} consists of five core components.
Central to 
\name{'s} automation capabilities are \textbf{flows} and their associated \textbf{rule-based triggers}. Flows are multi-step processes that describe an action to be performed. As previously mentioned, an ingestion flow might consist of three steps: 1) fetch the data, 2) preprocess the data, and 3) store the preprocessed data. When a flow is registered to \name{}, a rule-based trigger is defined to specify under what conditions the flow should be executed. 
%Rule-based triggers are conditionals that must be satisfied to trigger an automation. 
\name{} supports either periodic (i.e., automation occurs at a regular interval) or event-based (i.e., flows are reexecuted whenever their dependencies are updated) triggers.

% Privacy constraints and need for specialized hardware are necessities to many scientific applications. Whereas researchers may have direct access to this infrastructure, hosting these resources on the cloud is non-trivial. Therefore, 
\name{} adopts a novel distributed \textbf{compute and storage} model. Using this approach, \name{} is deployed at a globally accessible location (e.g., the cloud), while data and storage resources are hosted by the user. This approach allows resources to seamlessly scale with each new user and provides users with the flexibility to use \name{} on otherwise private data, as those data remain on their resources. \name{} will track the location of the data and the compute endpoints to ensure steps of the flow occur at the designated site with the correct inputs, but \name{} never obtains access to the data itself.

To maintain records of all automated computation and monitored data, \name{} includes a \textbf{metadata registry} that stores metadata pertaining to monitored data (e.g., where the data are stored, file format, how many different versions), data preprocessing and user-analysis metadata (e.g., which type of flow, which actions are executed, the rule that triggers the flow to execute) and  provenance-related metadata (i.e., what were the specific inputs and function used to generate the output data). This metadata facilitates data sharing of otherwise decentralized data, reproducibility, and replay of previously automated results. 

\name{} is designed with a comprehensive \textbf{security}
model in which user data remains on their resources, 
users delegate authorization to shared data and infrastructure
via a sophisticated OAuth-based model, and all data, flows, 
and compute functions are associated with authorization permissions.
% While much of \name{'s} \textbf{security} model relies on maintaining user data on user resources, delegating authorization to share data and infrastructure to the user, \name{} does provide security to even the metadata information. 
% At this moment in time, all users desiring to access \name{} must have a Globus account.
Further, only the users who registered the flows with \name{} can view flow progress. Like data and compute, the users who are authorized to monitor the running flows may extend authorization to other users.

% Once a rule is registered in \name{}, a trigger will be set that will launch a flow. The trigger may either be timer-based (i.e., will be executed at periodic intervals) or occur after \name{} receives a \textit{event}. An event in \name{} consists of any update operation, such as monitoring new data or \name{} flows generating new versions of monitored data. When a rule's conditions are met, a trigger will invoke the flow that will call upon a user's desired action.

% \name{} provides two predefined flows which describe the patterns seen in many scientific analyses: 1) a data ingestion flow and 2) a data analysis flow. The data ingestion flow is a three step flow responsible for fetching data located in a remote public repository external to \name{}, validating and/or transforming the data via user-defined code and updating the \name{}'s metadata. In contrast, the data analysis flow is a one or two-step flow that may execute a more complex user-defined policy as a first action and the user-define analysis as a second flow action. Should any action within either of either flows fail, the user may be notified of the outcome. 

% \valerie{This section will discuss the \name{} architecture. "what"}

\subsection*{\name{} Service and Client Library}
\name{} is implemented as a FastAPI~\cite{grinberg2018flask} web service with state maintained in a PostgreSQL database. To facilitate user interaction, a client command-line interface (CLI) and Python library is provided. The service is currently deployed on a virtual machine hosted by Argonne’s Computing, Environment, and Life Sciences (CELS) directorate and accessible through the swagger interface at \url{https://aero.cels.anl.gov/docs}. %virtual machine available at \valerie{insert URL here}.
% Users may interact with \name{} via HTTPS or the \name{} Python API and CLI.

The FastAPI service provides various functionalities to users. Through the service, users can look up stored metadata, retrieve monitored data, and register automated flows. The \name{} service uses Globus Search~\cite{ananthakrishnan2018globus} to provide flexible metadata querying capabilities. For data retrieval, the \name{} service provides the client with a reference to the Globus Collection on which the data are stored. 
%forwards to the client metadata which includes the Globus Collection on which the data is stored. 
The client uses the metadata to initiate a direct HTTPS GET request on the collection to retrieve the data. Ensuring data are only communicated with the client and never pass through the server. 
%ensures \name{} never obtains ownership of the data.

The client library provides a \texttt{register} function that wraps the user-provided action with the decorator \texttt{aero\_format} before registering it as a function with Globus Compute. The decorator is responsible for gathering the \name{} data, which are provided as inputs to the function, and storing them in a temporary directory, passing the filenames to the function. The function can then load the data from the file system, perform the desired operation, and return the output file name and some associated metadata, should the outputs be monitored. The wrapper then uses the output metadata to update \name{'s} state and stores the data in the desired Globus Collection. Listing~\ref{lst:noopfunc} presents an example of how a user function can be defined in \name{}. This listing complements Listing~\ref{lst:flow_registrations}, which presents an example of how to register the flow that is based on the defined user function.

\input{code/aero-mock-analysis}

\subsection*{Registering Automated Flows}

\begin{figure}
    \centering
    \includegraphics[width=1\linewidth]{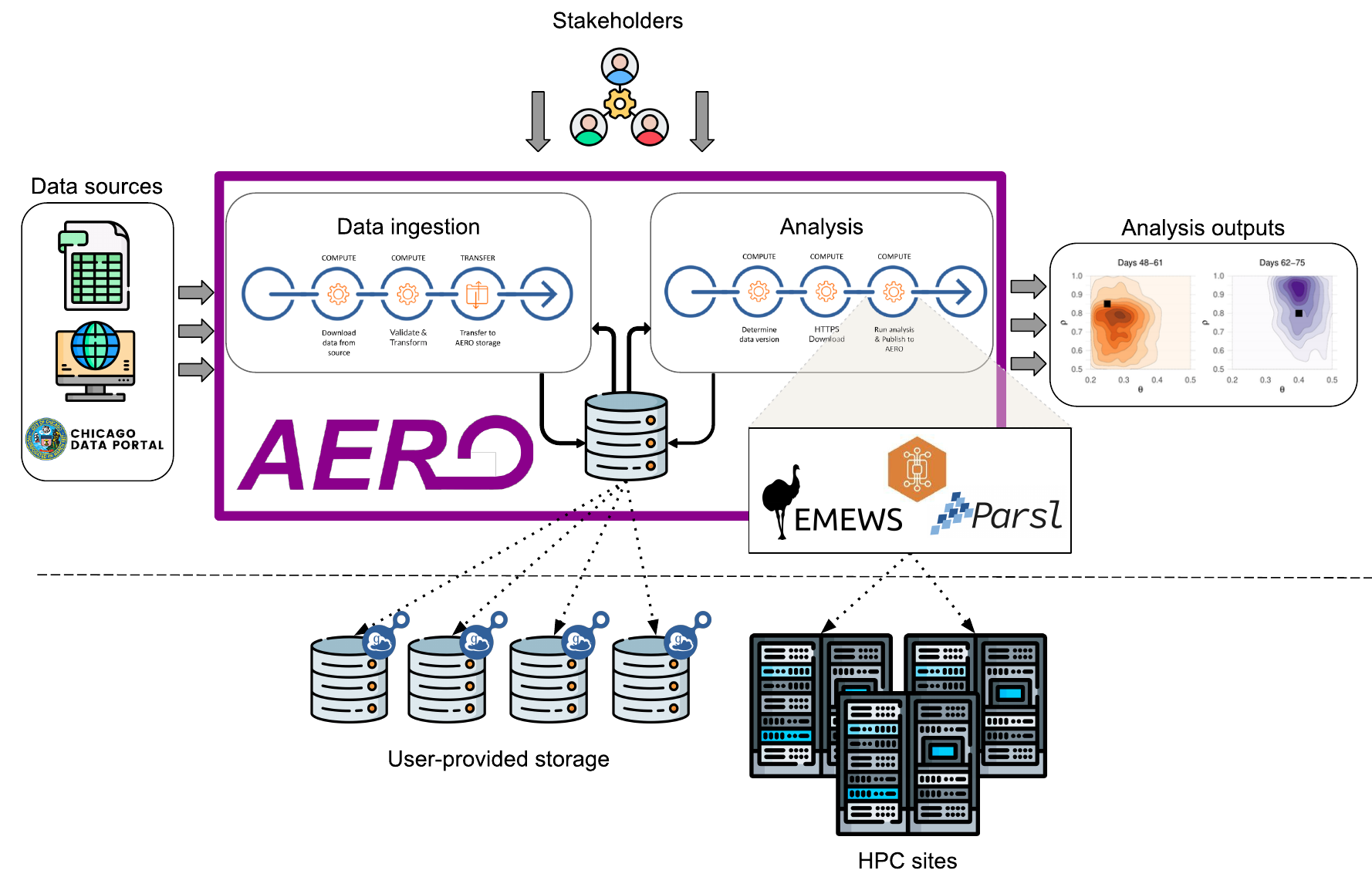}
    \caption{\name{} reference implementation event flow. Data is ingested from external sources via an automated timer-based ingestion flow that fetches, validates and stores the data on external storage at periodic intervals. The ingested data can then be queried through the centralized metadata server, and incorporated into automated analysis flows, which produce reports and visualizations that can then be shared with stakeholders, all through the platform. }
    \label{fig:flow-design}
\end{figure}

% \valerie{Requirements for registration}
% \name{} automates the execution of scientific applications by incorporating the applications within a flow. 
% \name{} enables automated execution of \textit{flows}---a sequence of steps, or \textit{actions}, that perform various operations. 

% \name{} supports two types of flows: 1) data ingestion and 2) analysis flows. 
% %, by allowing users to define the actions that compose them.  
% For \name{} to execute the automated flows, we require that users provide a Globus Compute~\cite{chard20funcx} endpoint UUID on which tasks will be executed.  Further details are provided in \autoref{lbl:byor}.

% In order to automate flow execution, a user must specify a rule for automation, any \name{} monitored inputs that are required by analysis, a Globus Compute endpoint UUID and the analysis function. 

% A user is required to provide various details about the flow when registering it in \name{}. 

\input{code/example}

% \valerie{flow execution}
% \valerie{need to introduce here GitHub Actions and present it as an alternative way to defining timers and flows in AERO}
Automated flows enable the continuous execution of repeatable scientific computations. These computations may include the ingestion of new data or the analysis of new data. A flow in \name{}, consists of a series of steps, which we refer to as \textit{actions}. These actions are responsible for acquiring input data from its stored location, performing any necessary computation (i.e., validation, transformation, or analysis), storing and gathering metadata of research outputs, and updating \name{'s} internal state. In \name{}, flows can be described using one of two different services: 1) Globus Flows or 2) GitHub Actions workflows. By allowing use of different frameworks, we enable users to describe flows using the services with which they are most familiar. 
%and harness the different functionalities made available by these two workflows.

When using Globus Flows, we adopt the Globus Compute Action Provider API to represent arbitrary asynchronous actions. Briefly, this API provides a common abstraction on top of external services for invoking and managing actions. Each action within a flow, with the exception of failure handling, uses the Globus Compute Action Provider to execute a Globus Compute function on the specified user resources. We use \globus{} Compute %was selected as the platform for executing functions 
for its ability to execute functions on a range of research cyberinfrastructure as well as its fine-grain authorization model and integration with Globus Flows via the 
%be deployed on the wide array of infrastructure available to researchers in addition to its integration with the 
Globus Compute Action Provider API. 
% Using the combination of the Globus Action Provider API and Globus Compute, users can execute individual functions within a flow across many different resources.

Workflows represented through GitHub Actions also rely on Globus Compute to execute remote tasks. \name{} GitHub Actions flows make use of the \texttt{CORRECT action}~\cite{correct}, a GitHub action enabling the execution of arbitrary functions on remote resources using Globus Compute. When using GitHub Actions, \name{} delegates all automation (i.e., trigger definition) to GitHub Actions, meaning \name{} primarily manages accounting in this more decentralized approach.

Two types of flows can be defined in \name{}: 1) ingestion flows and 2) analysis flows (Figure~\ref{fig:flow-design}). %Flows defined within \name{} are made up of two or three actions which are all executed on user-provided resources. 
Ingestion flows are made up of three actions, of which the first and last are defined by \name{}: 
1) retrieve data from remote location, 2) execute user provided validation and transformation functions, and 3) update \name{} database on success. Data are first downloaded from a remote source via an HTTP accessible path, thus necessitating that remote data locations provide a HTTP(S) API, and stored into a temporary location on the user's compute resources. The download action also extracts metadata from the data (e.g., file type, size, checksum) and verifies whether the checksum has changed since the last execution. If the data have changed, the user-provided verify and transform function is executed. On success of the preprocessing steps, metadata are stored by \name{} and the data are stored on user-provided storage.

If data ingestion, curation, and transformation complete successfully, the checksum of the data is computed and compared with data existing in \name{}. If the data already exists, no new data are added. Otherwise, the data are moved from the temporary directory and placed into a persistent folder located in the collection. \name{'s} state is then updated with all necessary metadata including the data source, name, and description, as well as how many versions are available, the path of the file on \gcs{}, and a timer interval for when to trigger fetching.

% Similarly, 
The analysis flow 
%is simpler compared to the ingestion flow. It
consists of three actions: 1) acquire the metadata on the specified version of the data, 2) execute the user-specified analysis, and 3) update the captured metadata. Like the ingestion flow, these actions all executed on user-provided resources via a Globus Compute endpoint. Unlike the ingestion flow, the user-provided analysis function downloads the data internally stored in \name{} in addition to registering any final results with \name{} for monitoring. Provenance information is recorded once the flow registers output data with \name{}. Listing~\ref{lst:flow_registrations} %~\ref{listing:reg-ingestion} and~\ref{listing:reg-analysis} 
demonstrates how an analysis flow can be registered with \name{}.

Failures are an inevitable part of running automated flows. For example, if data sources are offline, data formats change, or if remote computing resources are unavailable. Upon failure, Globus Flows will attempt to retry for a period of time. With terminal errors, for example, if the curation produces an error, an email is sent to the email address provided when the flow was registered with \name{}, informing the user of an error within the flow.
% Errors may occur within any action of a flow. By default, \name{} 
% %allows users to provide an email when registering a flow. This email will be used to
% can notify the user of any errors via email. 
% %that have occurred within an action of the flow. 
% If an error occurs within the user-defined code, the user will need to update their code and register a new flow with \name{}. For stochastic errors, like failed network connection due to temporary outage, the notified user may select to rerun the flow or wait until the next trigger for the flow to run.

% \valerie{add some figure here demonstrating the flow}

% \valerie{failure handling}

% \valerie{This is the user-side "how". Implementation "how" is in methods}

\subsection*{Automation triggers}
% \valerie{This is the when}
% \valerie{describe rules instead policies and move this to methods}

% \begin{table}[h]
%     \caption{\name{} rules and associated triggers} %\ian{ wonder if the word "policy" is right here. The first and the third seem to be actions that are triggered by an event: like an ITTT rule. I am not clear on the second. See the "Ian" text at start of section 3.2.}}
%     \centering
%     \begin{tabular}{ |p{6cm}|p{5cm}|  }
%         \hline
%         \textbf{Rule} & \textbf{Triggered by} \\
%         \hline
%         Time elapsed & Timer \\
%         \hline
%         All analyses inputs have been updated & Data version entry added to server \\
%         \hline
%         Some analyses inputs have been updated & Data version entry added to server \\
        
%         \hline
%     \end{tabular}
%     \label{tab:policies}
% \end{table}

% \ian{I wonder if it will be good to refer to ``trigger action programming'' here~\cite{ur2014practical}. That seems to be the model, and it is nicely described in that paper. Then the paper can talk about A) the three types of \textit{triggers} that are currently supported and B) the \textit{actions} that are supported (I am not yet clear on this); and C), as a policy is then precisely a trigger-action rule [or recipe, according to IFTTT terminology], give examples of policies that can be specified given available triggers and actions..}

The invocation of scientific flows is dependent on the occurrence of certain events. These events come in the form of timers, when the data sources are external to \name{}, and \name{} state updates for internally monitored data. For these reasons, we adopt a rule-based ``if-then'' model of trigger-action programming~\cite{ur2014practical}. We define a \textit{trigger} to be an event (e.g., time or update) which validates the conditions of a \textit{rule}. Should the conditions defined by a rule be met, a flow, be it ingestion or analysis, will be executed.

\name{} supports three types of rules and two different types of triggers. Two trigger types that result in rule evaluation are:  1) periodic and 2) state updates. For instance, all ingestion flows use timers, resulting in a rule condition being met whenever a specified period of time has elapsed. For the timers, 
we used the \globus{} Timers service to trigger the flows at the specified interval. The advantage of using \globus{} Timers for event triggering is that they are cloud-hosted.

In \name{} we use Globus Timers to invoke arbitrary action providers, such as compute functions and flows. When a flow is registered with \name{}, it specifies the rule-based trigger desired for that flow. Should the trigger be periodic, \name{} will initialize the timer trigger as it updates its state to include the new flow.

Analysis flows are more complex as they rely on updates to  their input parameters to determine whether they should be rerun. \name{} defines three rules for automation of analysis flows. Like for ingestion flows, the analysis flow can also use a time-based rule to define when the next run of the flow should execute. We imagine this to be used when input data is not monitored by \name{} and is ingested by the analysis action. More common rules, such as rerunning a flow when either some or all of the monitored action input data has changed are also defined by \name{}. 

There may be instances where policies for invoking flows need to consider more than just an update to input data. For instance, some analyses may require that values have changed above a certain threshold or that a particular value exists within the data. For these cases, as \name{} does not manage any data locally, an action containing a user-defined policy must be described by the user and executed alongside the analysis function.

When an analysis flow is invoked, other analysis flows that depend on the outputs may also be triggered. \name{} achieves this by maintaining a provenance tree of all functions and their dependencies. When an output is updated, \name{} determines if that output is an input dependency to other functions. If it is, \name{} will check to see which rules have been satisfied and run the flows belonging to those rules.

% Mechanisms to detect changes and trigger actions are essential for automation. \name{} is able to trigger events in the following scenarios: when timer-based delays are met, when new data are added to the database, and when certain conditions are satisfied by analysis outputs. Table~\ref{tab:policies} demonstrate the policies currently accepted by \name{}.

% Timer-based events are used to invoke data ingestion flows, and are optionally enabled for data analysis. For data ingestion, \name{} obtains data from external sources specified by users. These sources can range from public data provided by open data portals, to restricted data accessible by the service. Such external sources may be updated at regular intervals and would need to be captured by \name{}. For this reason, we allow users to specify a timer interval when adding new data source to trigger remote fetches of data updates. 

% As new data are ingested into the system, analyses may be triggered. Triggers occur when updates occur to the underlying metadata. These changes include updates to all analysis inputs, updates to any analysis inputs, or updates to specified inputs. We limit triggers to be set on changes in analysis inputs as we expect that changes in function inputs will be the only reason for re-execution, and in cases where that is not the case, a timer-based trigger can be set.

\subsection*{Distributed storage and compute model}\label{lbl:byor}

\begin{figure}
    \centering
    \includegraphics[width=0.5\linewidth]{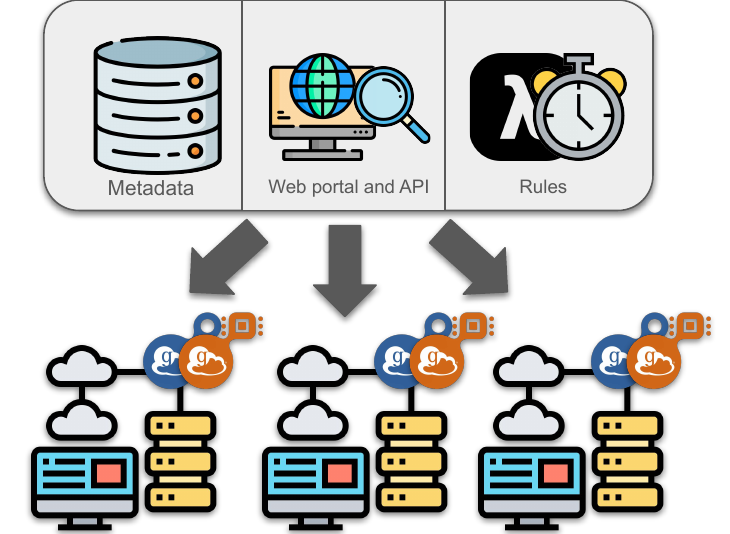}
    \caption{\name{'s} bring-your-own-resources model. Centralized services in \name{} include a database for capturing metadata, the web services, and the user-specified rule-based triggers}
    \label{fig:byor}
\end{figure}
Researchers have access to various computing resources including specialized computing resources and may need to moniter large volumes of data. For these reasons, \name{} relies on user-provided compute and storage resources (Figure~\ref{fig:byor}). Using this approach, \name{} can orchestrate actions across existing user-owned and -managed resources, reducing costs, increasing security, and enabling scaling  
both in compute and storage with increasing number of users. 
Moreover, users can pre-configure their resources as necessary to run their flows and can easily perform access control to their data. Furthermore, the decentralized distributed architecture relieves \name{} from requiring ownership of, or access to, user data.

To connect the external resources with \name{}, we use Globus~\cite{foster2011globus} data and compute services. 
%such as Globus Connect Server and Globus Compute.
We chose to rely on the Globus suite of services due to their ubiquity in scientific computing. Globus Connect Server allows users to easily and rapidly share and transfer data between distributed locations. To facilitate execution on a wide range of computing infrastructures, we use Globus Compute, a Python-based federated Function-as-a-Service~(FaaS) platform that can seamlessly run on HPC, cloud, and local workstations. \name{} requires each resource to have a Globus Connect Server (GCS) deployed and Globus Compute Endpoint. With GCS, users are able to manage access permissions to their data and provide high-performance third-party transfers between locations. Globus Compute, on the other hand, allows functions to be executed on remote computing resources. 
%users to incorporate their research infrastructure for optimal processing. 
%to enable data transfers to locations without Globus Connect Server configured. 

To monitor data within \name{}, users must provide a Globus Collection to which they have read and write permissions to store the data. Should users desire to share the data with other parties, they can modify the access permissions of their collections using the Globus Web application. 

Data monitored by \name{} relies on an object store-like approach for organization. \name{-monitored} data are stored in the \globus{} Collection following a flat hierarchy, replacing the data's original filenames with randomly-generated UUIDs. To obtain any details pertaining to the file's metadata, the \name{} service must be queried. 
%The use of a UUID over a more traditional filename aims to avoid issues that arise from duplicate filenames. 
\name{} maintains information such as storage location, user-specified filename, free-text description on the data provided by the user, file hash, and size, among other metadata. Users may use the provided client interfaces to obtain file-related metadata.

When flows are triggered, they create new versions of existing data. Only when data have been modified from their previous version are they stored in the \globus{} Collection and \name{'s} state is updated. %To determine whether there has been a modification between the previous and current version, a checksum of the current version is calculated and compared to the previous version. 

 % Globus Compute is not limited to running Python functions, and can be used as a wrapper to calling code written in any other language. By adopting Globus Compute, users can preconfigure their endpoints as required to successfully run their code.
 When configuring a flow to run on \name{}, users must register their functions with Globus Compute, and provide to \name{} the corresponding registered function's UUID and the Globus Compute Endpoint UUID for which to run the analyses.

% \subsection{Data management}

% Data within \name{} is entirely decentralized, with users registering their own \globus{} Collections for storage. We use this approach as it allows users to maintain ownership of their data and easily control access via the Globus SDK and web app. Furthermore, the use of \globus{} Collections allows users to utilize whichever storage is accessible to them, be it traditional filesystems (e.g., ext4, Lustre, DAOS), object stores (AWS S3, Google Cloud Storage), or web-hosted cloud storage solutions like Google Drive or DropBox as Globus provides a uniform API to arbitrary storage systems. To share data, \globus{} provides the ability to initiate transfers between two \globus{} endpoints, and alternatively, upload/download data directly via HTTPS, in instances where a secondary endpoint is not available. As \globus{} Connect Server is widely deployed on HPC, many users will have access to at least one \globus{} Collection on which they can store data.

\subsection*{Search}

\begin{figure}
    \centering
    \includegraphics[width=\linewidth]{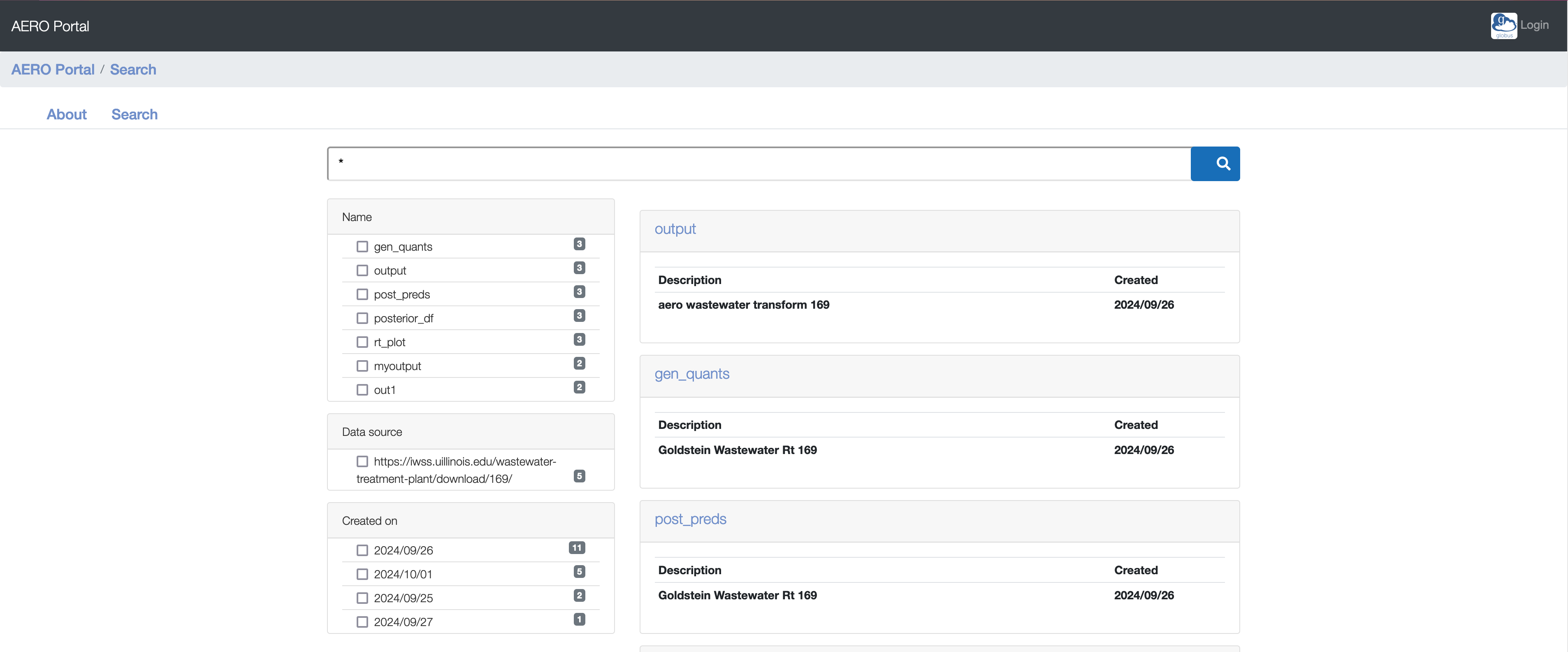}
    \caption{\name{} Search web interface. }
    \label{fig:search}
\end{figure}
Data in \name{} must be discoverable, with their corresponding ID and version easily accessible to users. \name{} leverages the Globus Search service to provide free-text search over managed data. Globus Search builds on ElasticSearch, provides an isolated and user-manageable cloud-hosted index, and implements a fine-grain permission-based model to control access to row-level metadata. 
%a thin wrapper around ElasticSearch and integrates nicely with other \globus{} services.

As new sources or data outputs are added to \name{}, entries are created and registered with \globus{} Search. The entries stored in the Search service are metadata consisting of: name, description, version, original source reference, URL to retrieve from \gcs{}, tags, data size, checksum, date created, and provenance. Users can interact with the search index using \name{}'s client API or web interface. An example of the web interface can be seen in Figure~\ref{fig:search}.
% Similar metadata is captured in output data entries, except they also include some provenance-related metadata that can guide users into reproducing the generated outputs.

\subsection*{Security Model}

It is critical that \name{} implement a robust security model as it manages public and private data, and executes functions on arbitrary user-specified resources. 
%, have some form of security enabled. 
All public-facing interfaces within \name{} perform authentication via the \globus{} Auth service. 

\name{'s} distributed compute and storage model integrates remote resources managed by users. We build on Globus Auth to ensure that entities are authenticated and that all actions are authorized. Globus Auth implements OAuth~2. \name{} is registered as a Globus Auth resource server and thus, users can authenticate with \name{} using any identity provider supported by Globus Auth, and \name{} can perform actions on behalf of users via a user-consented authorization model.

% All \globus{} services require authentication by default. The platform implementations ensure that the services within the platform can communicate with each other using shared credentials. However, all code executed on user-provided endpoints must be authenticated with the user's own credentials to run. Furthermore, users will only be able to access data which has been made available to them. Token-based authentication will be required for any action that requests to view or access the data. Moreover, any interaction with the client API will also require authentication, but all users with a \globus{} account are authorized to perform REST calls using the client interface.

% \subsection{Security model}

% \name{} is intended to be a data-sharing platform for automated analysis. It 

%Due to its decentralized design and ability to access user resources, it is critical that good security measures are employed.

% \name{} relies on  the security features provided by the underlying tools used to ensure security of the system as a whole.
% Workflows are managed by Globus Flows. 
As with all Globus Flows, each flow in \name{} is represented as its own Globus Auth resource server and thus is provided with credentials on behalf of the user. The flow requests consents for each of its actions (e.g., data movement, computation) to which the user must agree before the flow can be executed. Thus, actions in the flow can operate on the user's resources to access data and execute compute functions.

Data transfer occurs between the client and the desired Globus collection without any mediated interaction with \name{}. The user is asked to authenticate the first time they attempt to access a collection. The \name{} client then uses the generated tokens to initiate data transfer. The tokens are maintained locally and are never transferred to the \name{} server. 
%\name{} requires users to pre-authenticate with the Globus Collection required by the flow where the endpoint resides. The authentication tokens that authorized with the Globus Collection are then cached at the endpoint site. 
%When a flow executes, the actions executed within the flow rely on the locally-cached tokens to interact with the Collection. These tokens are never shared nor sent back to \name{}.
% \kyle{<ADD DETAILS OF HOW A FLOW USES DATA>}

Computation is performed on remote resources using Globus Compute. \name{} requires that a Globus Compute endpoint be deployed and that the \name{} Globus Auth client be permitted to execute functions on that endpoint. This can be accomplished by deploying a Globus Compute single user endpoint and enabling sharing~\cite{chard20funcx} or by deploying a multi-user endpoint and setting the identity mapping to map the client identity \cite{ananthakrishnan24multiuser} to a specific local user account. In both cases, additional security policies can be enforced, including white listing of functions that can be executed.

%% file: code/aero-mock-analysis.tex
\begin{lstlisting}[
    caption={%
        Example \name{} user function definition. \texttt{aero\_data} represents a path on the local file system generated by the \name{} function wrapper which retrieves the data from the Globus Collection and stores it locally. \texttt{output\_path}, uses the flow-specified output metadata to determine where the path should temporarily be stored. Other parameters are  maintained in the flow metadata, but are not otherwise monitored by \name{}. The function outputs an \texttt{AeroOutput} flagging to \name{} that the output should be monitored. The object contains the path to the stored file as well as a name, which is used by the wrapper to match output data to the flow-defined outputs and generate the provenance records.
    }, 
    language=Python,
    label={lst:noopfunc}, 
    float=t,
    boxpos=c,
    linewidth=0.985\columnwidth,
]
@aero_format
def wastewater_analysis(
    aero_data: Path | str,
    output_path: Path | str,
    arg1: list[int],
    arg2:int
) -> AeroOutput:
  import pandas as pd
  from aero_client.utils import AeroOutput

  df = pd.read_csv(aero_data)
  do_analysis(df, arg1, arg2)
  df.to_csv(output_path)

  return AeroOutput(name="wastewater_report", path=output_path)
\end{lstlisting}

%% file: code/example.tex
\begin{lstlisting}[
    caption={%
        Registering an analysis flow with \name{}. Input and output data that is monitored by \name{} is specified using dictionaries containing a key pointing to the function parameter names. Other, non-\name{} keyword arguments are specified within a separate dictionary.
    }, 
    language=Python,
    label={lst:flow_registrations}, 
    float=t,
    boxpos=c,
    linewidth=0.985\columnwidth,
]
from aero_client.api import register_flow
from aero_client.utils import PolicyEnum

aero_input_data_id: UUID
collection_uuid: UUID
collection_url: str

# update params as needed, keys need to match function param names
function_args = {"arg1": [1, 2, 3], "arg2": 2}  

 # need to update id post ingestion and key name needs to match parameter name in function
input_data = {"in_data": { "id": aero_input_data_id, "version": None }}

output_data = {"out_data": { 
    "collection_uuid": collection_uuid,
    "collection_url": collection_url
    }
}

description = "my analysis"
policy = PolicyEnum.ALL
fl = register_flow(
    endpoint_uuid="",
    function_uuid=function_uuid,
    input_data=input_data,
    output_data=output_data,
    kwargs=function_args,
    policy=policy,
    description=description,
)
\end{lstlisting}

%% file: sections/evaluation.tex
\section{Evaluation}

\subsection{Synthetic application}
% \valerie{Will move the use cases here}

We explore the functionality and performance of \name{} through the use of a synthetic application that executes many concurrent flows. The application executes an increasing number of concurrent no-op ingestion flows that read data from a Globus Collection, compile and store related file metadata in \name{}, and upload output data to a Globus Collection. 
%To demonstrate \name{'s} ability to scale with increasing traffic, 
We scale concurrency from 1 to 20 and compare the execution time between the two available execution models: Globus Flows and GitHub Actions. The ingested file used is 61.68~MB of randomly generated binary data. Synthetic application experiments were repeated 5 times.

As \name{} is configured to prevent registration of duplicate flows, we implement our user-defined validation/transformation to accept an unused random variable as a parameter.

%\valerie{Add a figure here of the flow for clarity?}

\subsubsection{Infrastructure}
We performed our experiments on the Chameleon Cloud testbed~\cite{keahey2020lessons}. The \name{} server was deployed on an m1.large KVM@TACC instance with 8~GB of RAM and 4~VCPUs. Compute instances used ``compute\_icelake\_r750'' bare-metal instances located at CHI@TACC. The compute instances, configured to run Globus Compute Endpoint version 2.34.0 on Python 3.10.16, contain 256~GiB of RAM and two Intel(R) Xeon(R) Platinum 8380 @ 2.30GHz CPUs. For the GitHub Actions workflows, we used the Ubuntu runners provided by GitHub. These runners invoked Globus Compute via a custom action.

The synthetic workflows input and output data were ingested from and stored in Argonne Leadership Computing Facility's (ALCF) Eagle file system, configured with Globus Connect Server (GCS).

\subsubsection{Results}
\begin{figure}[]
    \centering
    \includegraphics[width=0.7\linewidth]{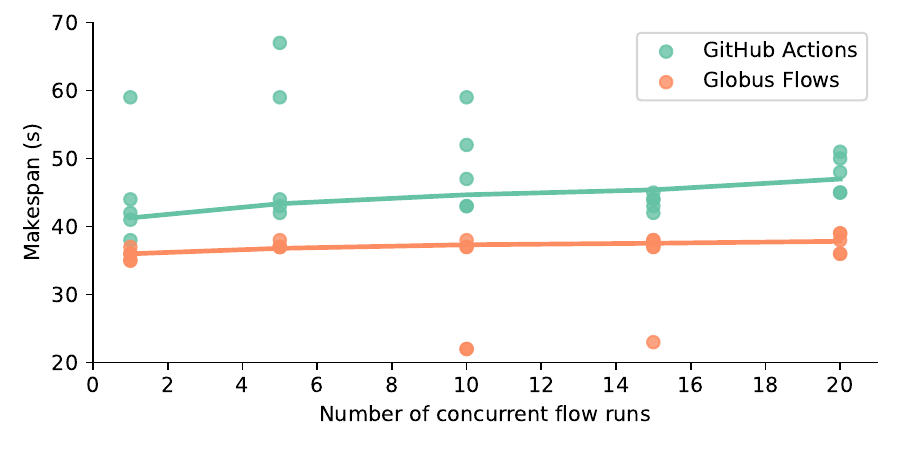}
    \caption{Makespan of synthetic ingestion application using GitHub Actions and Globus Flows over 5 repetitions}
    \label{fig:synthetic-makespan}
\end{figure}

\begin{figure*}[t!]
    \centering
    \captionsetup[subfigure]{justification=centering,labelformat=empty}
    \begin{subfigure}[c]{\textwidth}
        \centering
            \includegraphics[width=1\linewidth]{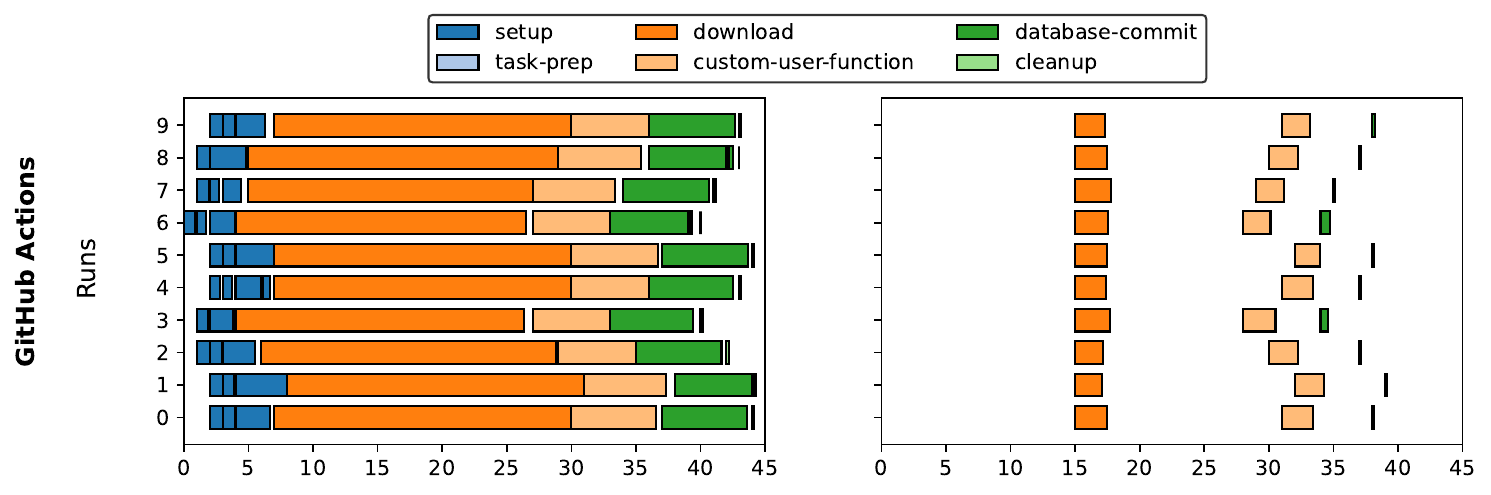}
        % \caption{1. GitHub Actions}
    \end{subfigure}\\%
    \begin{subfigure}[c]{\textwidth}
        \centering
            \includegraphics[width=1\linewidth]{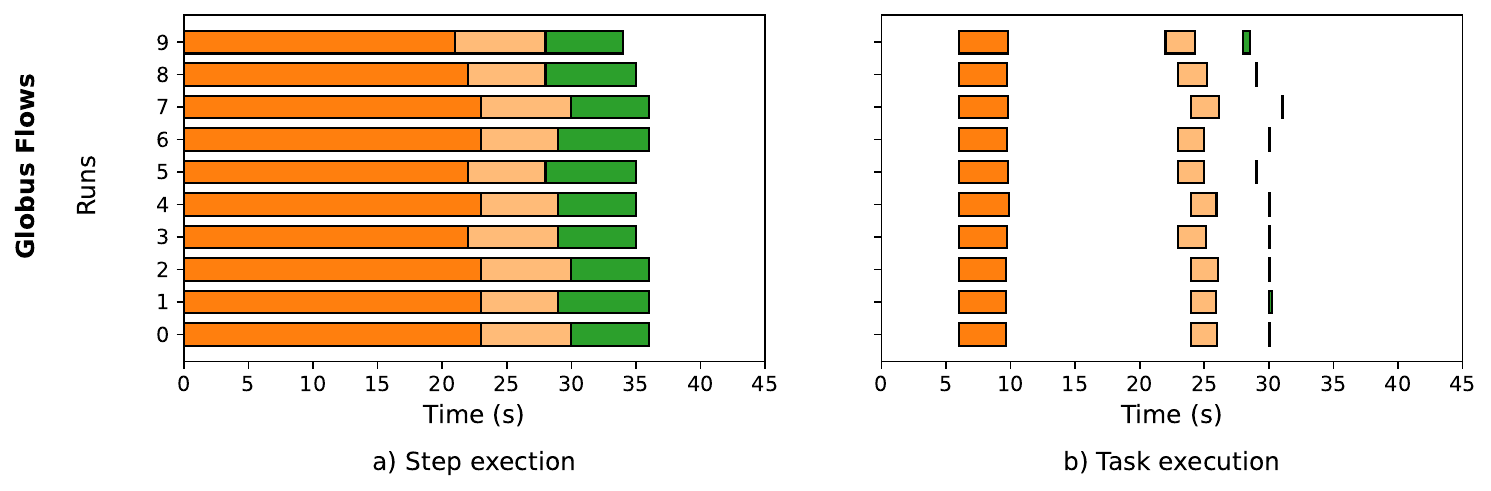}
        % \caption{2. Globus Flows}
    \end{subfigure}
    \caption{Synthetic application step and task execution. \textbf{Top row:} GitHub Actions step and task execution times. \textbf{Bottom row:} Globus Flows step and task execution times. \textit{Step} execution denotes the step start and end times reported by the framework. \textit{Task} execution denotes the start and end time reported by the tasks.}
    \label{fig:gantt-40}
\end{figure*}

% \begin{figure*}[t!]
%     \centering
%     \captionsetup[subfigure]{justification=centering}
%     \begin{subfigure}[t]{0.5\textwidth}
%         \centering
%             \includegraphics[width=\linewidth]{figures/gf-rep_4-histogram-Real.pdf}
%         \caption{Globus Flows}
%     \end{subfigure}%
%     \begin{subfigure}[t]{0.5\textwidth}
%         \centering
%        \includegraphics[width=\linewidth]{figures/ga-figures-histogram-Real.pdf}
%         \caption{GitHub Actions}
%     \end{subfigure}
%     \caption{Histogram of synthetic application task execution time. Task times reported represent actual task duration.\valerie{might need to normalize y-axis between actions and flows}}
%     \label{fig:histogram-40}
% \end{figure*} 
% % scale the data experiments

Scaling experiments using our synthetic application demonstrate that \name{} exhibits near-perfect scaling with increasing load (\autoref{fig:synthetic-makespan}). While variability was high between the individual runs, further analysis indicated that variability was a result of relying on production services (i.e., GitHub Actions, Globus Flows, and Globus Compute). For example, with GitHub Actions, it sometimes occurred that the runners were not all provisioned at the same time, resulting in longer makespans. Higher-than-expected load on Globus services led to additional delays in workflow processing. However, neither GitHub Actions nor Globus Flows make performance guarantees at the granularity of tens of seconds.

Our results in \autoref{fig:synthetic-makespan} also showed that Globus Flows is on average faster than GitHub Actions. 
%across the scaling experiments.
Individual repetitions (\autoref{fig:gantt-40}) showed that the difference can be explained from the fact that GitHub Actions has additional actions that take place before and after \name{-related} step execution to setup and clean up the runner.

Comparing task execution times with step times in \autoref{fig:gantt-40}, we can see that steps report longer durations than the actual task duration. This is due to the exponential decay in polling interval implemented by the Globus Compute Action Provider and the custom GitHub Action to determine when tasks have completed without overloading the service. Nevertheless, \name{} aims to provide automation and does not guarantee low latency.

% As can be seen in ~\autoref{fig:synthetic-makespan}, while execution time of GitHub Actions flows varies, on average, the execution times of GitHub Actions and Globus Flows are similar. Overheads in both GitHub actions and Globus Flows Looking more closely at task execution ~\autoref{fig:gantt-40}, we can see that task start times and round-trip time of tasks can vary greatly between the two execution frameworks. For instance, we notice that for Globus Flows, the start times of the tasks seemingly occur in batches. The reasoning for this 

% \begin{mybox}{Use case: $R(t)$ estimation for continuous monitoring of ongoing epidemics}
% \begin{Frame}[
\subsection{Use case: $R(t)$ estimation for continuous monitoring of ongoing epidemics}

As a motivating example, we demonstrate how \name{} 
% workflows that
automatically estimates the SARS-CoV-2 effective reproductive number, $R(t)$, based on published data. $R(t)$ is a time-varying quantity that represents, on average, the number of new cases caused by an already-infected individual throughout the span of their illness. It is a useful epidemic quantity for detecting trends in community disease transmission and informing policy interventions, and it is closely monitored by public health officials throughout the lifespan of an epidemic. $R(t)$ can be estimated using a variety of methods and data sources \cite{gostic_practical_2020}, all of which involve repeatedly (1) downloading, processing, and storing data, (2) running the analysis that estimates $R(t)$, and (3) reporting results \cite{huisman_estimation_2022}. With \name{}, we streamline this process, eliminate critical points of failure, and enable extensions of analyses to those with more computational demands, as we show below.

\vspace{1ex}
% Story is that the way you estimate Rt depends on what kind of data you have, which depends on what is available during different phases of pandemic. We demonstrate use-cases from two distinct phases of an epidemic, one with actively-reported data (clinical case surveillance), the other with passive surveillance datasets.
Infection events cannot be observed directly in a population, so proxy datasets are used to estimate $R(t)$ under reasonable assumptions about their relationship to infection incidence\cite{gostic_practical_2020}. While an epidemic is under active monitoring, governments may enact reporting mandates that result in high quality and frequently updated surveillance datasets, such as reported cases or hospitalizations. Simple analytical tools can be used to estimate $R(t)$ from these datasets, which we demonstrate in the first use-case. However, during other phases of an epidemic, reporting mandates may end, and datasets requiring opt-in may no longer be consistently updated. Our second use-case demonstrates a workflow that estimates $R(t)$ using a passive surveillance indicator (wastewater) and requires more complicated processing and analysis.

\vspace{1ex}

\noindent
\textbf{Clinical surveillance data:} % Active surveillance?
For much of the COVID-19 pandemic, the Chicago Department of Public Health (CDPH) published detailed surveillance datasets, including daily hospitalizations and emergency department (ED) visits associated with COVID-19 and COVID-like illnesses. As our first \name{} use-case, we adapt a study by the CDPH and collaborators \cite{richardson_tracking_2022} that estimates $R(t)$ from these datasets following the methodology proposed by Huisman et al.\cite{huisman_estimation_2022}, implemented in the Python package epyestim v0.1\cite{epyestim}. Given a time series of infection incidence, the Cori method~\cite{cori_new_2013} is recommended for estimating  $R(t)$, as it only requires the incidence data and an assumed distribution of the time between primary and secondary infections (known as the generation interval) as inputs\cite{gostic_practical_2020}. To infer infection incidence from proxy observations like hospitalizations or ED visits, the observed time series are smoothed and deconvolved according to the delay distribution between infection and recorded event (e.g., how many days we expect between infection and hospital admission) \cite{goldstein_reconstructing_2009}. Hence, a pipeline for real-time $R(t)$ estimation from surveillance data involves three primary inputs: the time series, a distribution for the generation interval, and the delay distribution. Once these quantities are specified, the time series is transformed into an estimate of infection incidence, which is used to estimate $R(t)$ with the Cori method. 
% With \name{}, we developed a workflow that, given a source of clinical surveillance data and user-specified input distributions, estimates $R(t)$ with uncertainty and generates a report.

\vspace{1ex}

% At user-specified intervals, \name{} retrieves the hospital admissions and ED visits datasets from the Chicago Data Portal, storing them as \name{} data sources. Cori requires an observation for every time step, so an \name{} transformation is registered to fill missing values and return a clean time series. Prior to execution, the user specifies parameters of the two input distributions in a configuration file, which is ingested and stored by \name{}. A Python function, also registered with \name{}, generates a discretized version of the input distributions that are then deconvolved with the surveillance time series to infer the infection incidence, and then estimates $R(t)$. Finally, a report is prepared that records the $R(t)$ estimate along with its associated 95\% confidence intervals, saved with \name{}, and made available for viewing by authorized users.  

\vspace{1ex}

%\vspace{1ex}

%\noindent
%With the advent of COVID-19, wastewater surveillance has become an important data source for epidemiological modeling. Using measurements of the concentrations of pathogen genomes (e.g., the SARS-CoV-2 virus) in wastewater, public health officials and researchers can more quickly determine the trajectory of an epidemic, leading to more effective decision making. 
% Our wastewater use case leverages the work done by Goldstein et al.\cite{goldstein_semiparametric_2023} where they develop a model to estimate $R(t)$ from wastewater data. The model is a compartmental EIRR model with separate recovery compartments for those individuals that are recovered but still shedding viral RNA into wastewater streams, and those recovered individuals that are not. This compartmental model coupled with Bayesian non-parametric priors was shown to be successful at estimating $R(t)$ from pathogen genome concentrations in wastewater for both simulated and real data.

\noindent\textbf{Pathogen concentrations in wastewater:} 
The mandates that ensured consistent access to updated COVID-19 surveillance datasets have ended, and most of the datasets described in the previous use-case are no longer actively maintained. However, COVID-19 remains a public health risk and monitoring $R(t)$ in the absence of reporting mandates is nevertheless essential. Passive surveillance indicators, such as pathogen concentrations in wastewater, do not require populations to opt-in and can always be monitored. However, the signal from such data streams is noisy and subject to complicated dynamics, hence $R(t)$ cannot be as reliably estimated using the Cori procedure outlined above. As our second use-case, we leverage a compartmental model-based $R(t)$ estimation framework~\cite{goldstein_semiparametric_2023} (Goldstein method). This method combines a mechanistic epidemiological model (with separate recovery compartments for those individuals that are recovered but still shedding viral RNA into wastewater streams, and those recovered individuals that are not) and a separate statistical model of the observed pathogen genome concentrations in wastewater. $R(t)$ is estimated as a posterior distribution using a semi-parametric Bayesian sampling framework. Note that this estimation procedure is significantly more computationally expensive than the Cori method.

\begin{figure}[h!]
    \centering
\includegraphics[width=\linewidth]{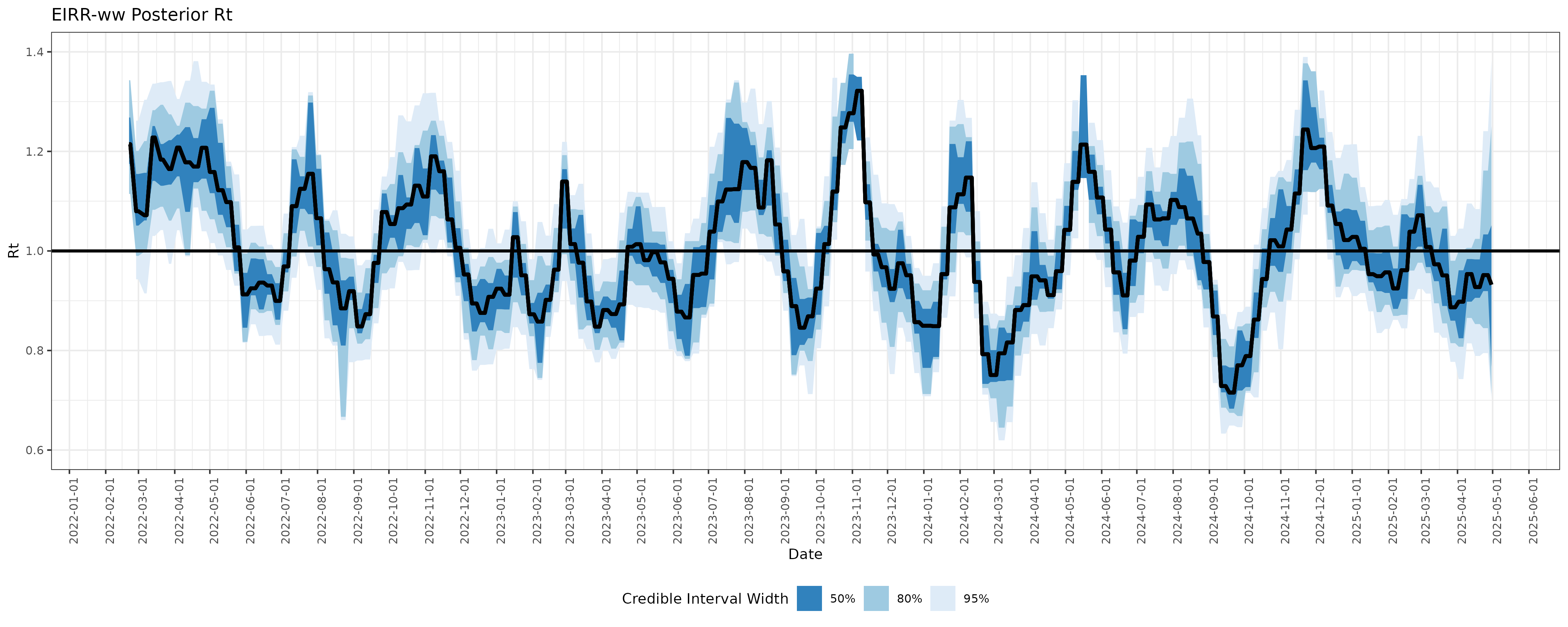}
\captionof{figure}{Example wastewater output report generated by \name{} registered analysis.}\label{fig:reff}
\end{figure}

% \vspace{1ex}
The Goldstein method was developed to estimate $R(t)$ from pathogen RNA concentrations in Los Angeles, CA, wastewater data~\cite{goldstein_semiparametric_2023}. 
We adapted their manual process into an automated 
\name{} workflow using Chicago wastewater data from the O'Brien water reclamation plant. This workflow consists of two parts: a preprocessing step, and the running of the model itself. The O'Brien water data requires a preprocessing step that transforms the raw data, added as an \name{} data source, into model input. The \name{} framework automates this such that a transformation function is run against the raw wastewater data whenever it is polled at the user specified interval. This transformed data can then be used in the $R(t)$ estimation model. The execution of the estimation model is implemented as a Python code harness that 1) retrieves and stages the wastewater data using the \name{} Python API; 2) executes a Julia code estimation model; 3) executes R code to create the $R(t)$ plots from the tabular data produced by the estimation model; and 4) stores the tabular data and plots to an \name{} collection, again using the \name{} Python API. Figure~\ref{fig:reff} displays one of the plots generated via \name{'s} automated analysis.

%% file: sections/discussion.tex
\section*{Discussion}
% \name{} %has been developed as a prototype for automated scientific research. \name{} achieves its goal by bridging the \globus{} 
% demonstrates the benefits of automated scientific research. 

\name{} addresses the requirements for automated scientific research with 1) rule-based triggers and automated flow execution on distributed, user-supplied compute and storage resources; 2) metadata extraction and provenance capture to facilitate sharing, discovery, and reproducibility; and 3) a flexible security model to control access to data and compute endpoints. We demonstrated \name{'s} capabilities using a synthetic workflow and showed its application in the public health domain by automating two COVID-19 epidemic modeling flows.
%These example applications tackled challenges that have long been recognized in public health emergencies~\cite{brown2015computer} and employed common analysis methods~\cite{siettos2015modeling}, while opening up the possibilities of more responsive decision support~\cite{xiong2021establishment,lima_value_2024}.
While we demonstrate the applicability of \name{} to epidemiological modeling, our model of fetch, compute, and publish, used in both ingestion and analysis flows, is generalizable to many domains. 
% We plan to expand on this model to support triggers based on data contents.
% More intricate sets of rules, such as those relating to triggering automation based on input data values will need to be added 

In database domains, extract, transform, and load (ETL) procedures are %necessary in all areas of data analysis and can 
often fully automated~\cite{vassiliadis2009survey}. Cloud platforms provide services, such as AWS Glue~\cite{Saxena2023} and Azure Data Factory~\cite{rawat2019introduction}, to automate ETL procedures. While these services automate ETL and subsequent analysis, they are restricted to their platforms and pricing models. Further, these services 
%provide out-of-the-box solutions, they 
are non-trivial to adapt to scientific computing contexts where a variety of heterogeneous hardware (e.g., workstations, HPC clusters, and cloud platforms) may be used in producing needed research outputs. In contrast, \name{} works with existing cyberinfrastructure, combining a cloud-hosted automation platform with an ecosystem of scientific computing resources.

Automation in science has largely been incorporated in the context of reproducibility. Live papers~\cite{brinckman2019computing,appukuttan2023ebrains, ellerm22live} seek to enable reproducible research by facilitating re-execution of experiments through the recreation of the environments used to obtain the results. More recently, Continuous Integration (CI) has been employed as a tool for automation and reproducibility. For example, NeuroCI~\cite{sanz2022neuroci} uses CI to automate re-execution of code in the context of ensuring reproducibility in neuroimaging contexts. Like \name{}, it enables users to set timer- and metadata-based (i.e., updates to git-tracked data and code) triggers. While this approach meets the requirements for reproducible research, it is bound to domain-specific solutions for the distribution and exploration of data. More broadly, Git-based solutions lack a unified API for querying all available metadata, hindering ease of collaboration across projects. 

% \begin{comment}
% \valerie{fix this to expand on shortcomings of CI and why we need a kind of system like \name{}}
% \end{comment}

% to run CI workflows automatically whenever data or code has been updated~\cite{sanz2022neuroci,hubverse,flusight}. Using

% More recently, GitHub and CI have been applied more broadly for data sharing and automation, as in the case of Huberse~\cite{hubverse} and Flusight~\cite{flusight}, however, the ability to automate analysis execution on various heterogeneous hardware and easily search the available data has not been streamlined. 

Lack of code and data sharing was highlighted as one of the main challenges during the COVID-19 pandemic. % that resulted in duplicated efforts. 
Curated software repositories, such as the Canadian Open Neuroscience Platform~\cite{poline2023data}, Sage App Catalogue~\cite{sage}, Pegasus Hub~\cite{pegasushub}, HuggingFace~\cite{huggingface}, and Kaggle~\cite{kaggle}, enable researchers to peruse available software prior to developing their own and in some cases provide a common approach to executing the available tools. These solutions, however, do not provide any automation nor do they facilitate deployment at different sites.

%\valerie{sage, waggle and deriva here}
As with code repositories, there are many data sharing platforms for use in scientific research. These platforms tend to be domain-specific (e.g., OpenNeuro~\cite{markiewicz2021openneuro} and The International Genome Sample Resource~\cite{clarke2017international}), governmental open data portal (e.g., Chicago Open Data Portal~\cite{cdp}), or generalized portals that catalog data from many decentralized sources (e.g., DataLad~\cite{wagner2022fairly,halchenko2021datalad}). Such data portals ensure FAIR sharing of data via unique identifiers, well-described metadata, capturing provenance of shared data, and providing interfaces for data discovery. \name{} embraces the FAIR principles for making data accessible to users; however, it goes further in automating the entire data-oriented research process and doing so in a way in which those automation flows are themselves FAIR.
\begin{comment}
  \valerie{don't know if this is a good argument, but basically a Globus provides auth and with the metadata server, you get it all without having to learn much}  
\end{comment}
% While these repositories facilitate the sharing of data, they do not provide mechanisms to provide privacy on their shared data, rather, they redirect to authorization methods available at the storage sites. This may prevent smaller entities from choosing to share their data due to overheads in ensuring privacy. 

Effective collaboration is often hindered by inherent difficulties in securely sharing data and code and executing code in a reproducible manner across different facilities. Various platforms have emerged to address these challenges. CBRAIN~\cite{sherif2014cbrain} is a web-based platform and RESTful API that provides data sharing, visualization, security, and facilitates analysis execution cross-sites. Tapis~\cite{stubbs2021tapis} is a RESTful API that, like CBRAIN, provides data sharing, security, provenance and distributed cross-site code execution capabilities. Additionally, Tapis provides capabilities to define automation on event-streams. This automation is achieved through Actor-Based Containers (ABACO) which can receive events and deploy containers for which the events will execute. However, Tapis relies on Docker and Kubernetes for deployment of functions, and therefore can only be used at computing sites pre-configured with Tapis. Rather than reinvent these capabilities, \name{} builds on the widely used Globus platform to enable data and compute management across systems.

Rule-based trigger action programming has been used in many contexts, particularly in the context of data management. The Integrated Rule-Oriented Data System (iRODS)~\cite{hedges2007management} is a data-management software that focuses on managing very large datasets. iRODS provides a Rule Engine to enable users to describe rules using their own domain-specific programming language and provides plugins to describe rules using various programming languages. Rules described in iRODS are commonly used for data storage and metadata operations. These include migrating files between different file systems, staging data, setting access-control permissions, modifying file formats, or transforming data. %Similarly to iRODS, 
Ripple~\cite{chard2017ripple} provides recipes for specifying rules that trigger actions on filesystem events. Ripple differs from iRODS primarily in that the data can reside anywhere and need not be cataloged by the system, as is the case with iRODS. In contrast to these engines, \name{} defines rules on \name{} metadata or periodic events. The rules enabled by the system are statically defined, restricting users to define additional policies within their automated flows. Future work will expand rule definition in \name{} to enable more flexibility.
\begin{comment}
\valerie{honestly, simpler is probably better, at least now anyway. I'm also wondering if AERO should maybe use these systems, because we have no proper handling of data movement. This goes with Dan Katz's comment on my presentation that if the URL changes, we can never locate the data}.
\end{comment}

Creating a collaborative platform for automated analyses comes with many challenges. As exemplified by all the tools that provide aspects of collaborative work or automation, many components exist, but integrating them to address use cases represents a complex undertaking. We developed \name{} using various Globus services, as they are commonly adopted and available in HPC settings, designed to be built upon, and provide a comprehensive authentication and authorization model. In an attempt to reduce the overheads of adopting new tools, we provide users with a simple Python API, which enables registration of functions and their respective flows, in addition to the ability to query and retrieve data. Querying and visualization of the data is also possible through a web-based interface. 
%While this work is currently in its early-stages, we hope to provide additional features, such as the ability for users to set notifications (e.g., ``watch'') on external flows, in the future.

% \valerie{talk about lessons learned}

% Following the 

% Recent efforts in COVID-19 pandemic modeling, such as Huberse~\cite{hubverse} and Flusight~\cite{flusight} \valerie{double-check} have adopted this style of approach, 

% To bind all the \globus{} services and provide additional functionality such as provenance, \name{} manages a metadata database that tracks data and the history of \textit{action} execution. Furthermore, \name{} enables description of policies beyond just timers, allowing policies to be written on database updates and also policies described within arbitrary Python functions.

% \valerie{adding related work here. listing related-work items}

% \begin{itemize}
%     \item things that do the same thing for cloud only
%     \item scientific efforts that are loosely related (CI for reproducibility, live papers, workflow management systems)
%     \item data sharing platforms
% \end{itemize}

%\input{sections/related_work}

% We have demonstrated \name{'s} use on 

%% file: sections/acknowledgements.tex
\section*{Acknowledgments}

This material is based upon work supported by the National Science Foundation under Grant 2200234, the U.S. Department of Energy, Office of Science, under contract number DE-AC02-06CH11357 and the Bio-preparedness Research Virtual Environment (BRaVE) initiative. This research was completed with resources provided by the Research Computing Center at the University of Chicago, the Chameleon testbed supported by the National Science Foundation, the Laboratory Computing Resource Center at Argonne National Laboratory, and the Argonne Leadership Computing Facility, which is a DOE Office of Science User Facility.

%% file: main.bbl
\begin{thebibliography}{10}

\bibitem{zhao2005notation}
Zhao Y, Dobson J, Foster I, Moreau L, Wilde M.
\newblock A notation and system for expressing and executing cleanly typed workflows on messy scientific data.
\newblock ACM Sigmod Record. 2005;34(3):37--43.

\bibitem{wilde2009parallel}
Wilde M, Foster I, Iskra K, Beckman P, Zhang Z, Espinosa A, et~al.
\newblock Parallel scripting for applications at the petascale and beyond.
\newblock Computer. 2009;42(11):50--60.

\bibitem{babuji2019parsl}
Babuji Y, Woodard A, Li Z, Katz DS, Clifford B, Kumar R, et~al.
\newblock {Parsl: Pervasive parallel programming in Python}.
\newblock In: Proceedings of the 28th International Symposium on High-Performance Parallel and Distributed Computing; 2019. p. 25--36.

\bibitem{deelman2015pegasus}
Deelman E, Vahi K, Juve G, Rynge M, Callaghan S, Maechling PJ, et~al.
\newblock {Pegasus, a workflow management system for science automation}.
\newblock Future Generation Computer Systems. 2015;46:17--35.

\bibitem{di2017nextflow}
Di~Tommaso P, Chatzou M, Floden EW, Barja PP, Palumbo E, Notredame C.
\newblock {Nextflow enables reproducible computational workflows}.
\newblock Nature biotechnology. 2017;35(4):316--319.

\bibitem{ozik_desktop_2016}
Ozik J, Collier NT, Wozniak JM, Spagnuolo C.
\newblock From Desktop to Large-Scale Model Exploration with {Swift/T}.
\newblock In: 2016 {Winter} {Simulation} {Conference} ({WSC}); 2016. p. 206--220.

\bibitem{collier_distributed_2024}
Collier N, Wozniak JM, Fadikar A, Stevens A, Ozik J.
\newblock Distributed {Model} {Exploration} with {EMEWS}.
\newblock In: 2024 {Winter} {Simulation} {Conference} ({WSC}). Orlando, FL, USA: IEEE; 2024. p. 72--86.
\newblock Available from: \url{https://ieeexplore.ieee.org/document/10838848/}.

\bibitem{vescovi22linking}
Vescovi R, Chard R, Saint ND, Blaiszik B, Pruyne J, Bicer T, et~al.
\newblock Linking scientific instruments and computation: Patterns, technologies, and experiences.
\newblock Patterns. 2022;3(10):100606.
\newblock doi:{https://doi.org/10.1016/j.patter.2022.100606}.

\bibitem{chard23automation}
Chard R, Pruyne J, McKee K, Bryan J, Raumann B, Ananthakrishnan R, et~al.
\newblock Globus automation services: Research process automation across the space–time continuum.
\newblock Future Generation Computer Systems. 2023;142:393--409.
\newblock doi:{https://doi.org/10.1016/j.future.2023.01.010}.

\bibitem{hubverse}
{Consortium of Infectious Disease Modeling Hubs}. {The hubverse: open tools for collaborative modeling}; 2023.
\newblock Available from: \url{https://hubdocs.readthedocs.io/en/latest/}.

\bibitem{flusight}
{Centers for Disease Control and Prevention, National Center for Immunization and Respiratory Diseases (NCIRD)}. {FluSight}: Forecasting for Influenza Prevention and Control; 2023.
\newblock Available from: \url{https://hubdocs.readthedocs.io/en/latest/}.

\bibitem{chard14globus}
Chard K, Tuecke S, Foster I.
\newblock Efficient and Secure Transfer, Synchronization, and Sharing of Big Data.
\newblock IEEE Cloud Computing. 2014;1(3):46--55.
\newblock doi:{10.1109/MCC.2014.52}.

\bibitem{leisman_modeling_2024}
Leisman KP, Owen C, Warns MM, Tiwari A, Bian GZ, Owens SM, et~al.
\newblock A modeling pipeline to relate municipal wastewater surveillance and regional public health data.
\newblock Water Research. 2024;252:121178.
\newblock doi:{10.1016/j.watres.2024.121178}.

\bibitem{runge_modeling_2022}
Runge M, Richardson RAK, Clay PA, Bell A, Holden TM, Singam M, et~al.
\newblock Modeling robust {COVID}-19 intensive care unit occupancy thresholds for imposing mitigation to prevent exceeding capacities.
\newblock PLOS Global Public Health. 2022;2(5):e0000308.
\newblock doi:{10.1371/journal.pgph.0000308}.

\bibitem{lima_value_2024}
Lima PNd, Karr S, Lim JZ, Vardavas R, Roberts D, Kessler A, et~al.
\newblock The {Value} of {Environmental} {Surveillance} for {Pandemic} {Response}.
\newblock Santa Monica, CA: RAND Corporation; 2024.

\bibitem{ozik_population_2021}
Ozik J, Wozniak JM, Collier N, Macal CM, Binois M.
\newblock A population data-driven workflow for {COVID}-19 modeling and learning.
\newblock The International Journal of High Performance Computing Applications. 2021;35(5):483--499.
\newblock doi:{10.1177/10943420211035164}.

\bibitem{hotton_impact_2022}
Hotton AL, Ozik J, Kaligotla C, Collier N, Stevens A, Khanna AS, et~al.
\newblock Impact of changes in protective behaviors and out-of-household activities by age on {COVID}-19 transmission and hospitalization in {Chicago}, {Illinois}.
\newblock Annals of Epidemiology. 2022; p. S1047279722001053.
\newblock doi:{10.1016/j.annepidem.2022.06.005}.

\bibitem{chang_mobility_2021}
Chang S, Pierson E, Koh PW, Gerardin J, Redbird B, Grusky D, et~al.
\newblock Mobility network models of {COVID}-19 explain inequities and inform reopening.
\newblock Nature. 2021;589(7840):82--87.
\newblock doi:{10.1038/s41586-020-2923-3}.

\bibitem{shea_multiple_2023}
Shea K, Borchering RK, Probert WJM, Howerton E, Bogich TL, Li SL, et~al.
\newblock Multiple models for outbreak decision support in the face of uncertainty.
\newblock Proceedings of the National Academy of Sciences. 2023;120(18):e2207537120.
\newblock doi:{10.1073/pnas.2207537120}.

\bibitem{bollyky_assessing_2023}
Bollyky TJ, Castro E, Aravkin AY, Bhangdia K, Dalos J, Hulland EN, et~al.
\newblock Assessing {COVID}-19 pandemic policies and behaviours and their economic and educational trade-offs across {US} states from {Jan} 1, 2020, to {July} 31, 2022: an observational analysis.
\newblock The Lancet. 2023;401(10385):1341--1360.
\newblock doi:{10.1016/S0140-6736(23)00461-0}.

\bibitem{watson_global_2022}
Watson OJ, Barnsley G, Toor J, Hogan AB, Winskill P, Ghani AC.
\newblock Global impact of the first year of {COVID}-19 vaccination: a mathematical modelling study.
\newblock The Lancet Infectious Diseases. 2022;22(9):1293--1302.
\newblock doi:{10.1016/S1473-3099(22)00320-6}.

\bibitem{ray_ensemble_2020}
Ray EL, Wattanachit N, Niemi J, Kanji AH, House K, Cramer EY, et~al.
\newblock Ensemble {Forecasts} of {Coronavirus} {Disease} 2019 ({COVID}-19) in the {U}.{S}.
\newblock Epidemiology; 2020.
\newblock Available from: \url{http://medrxiv.org/lookup/doi/10.1101/2020.08.19.20177493}.

\bibitem{borchering_modeling_2021}
Borchering RK, Viboud C, Howerton E, Smith CP, Truelove S, Runge MC, et~al.
\newblock Modeling of {Future} {COVID}-19 {Cases}, {Hospitalizations}, and {Deaths}, by {Vaccination} {Rates} and {Nonpharmaceutical} {Intervention} {Scenarios} — {United} {States}, {April}–{September} 2021.
\newblock MMWR Morbidity and Mortality Weekly Report. 2021;70(19):719--724.
\newblock doi:{10.15585/mmwr.mm7019e3}.

\bibitem{lima_reopening_2021}
Lima PNd, Lempert R, Vardavas R, Baker L, Ringel J, Rutter CM, et~al.
\newblock Reopening {California}: {Seeking} robust, non-dominated {COVID}-19 exit strategies.
\newblock PLOS ONE. 2021;16(10):e0259166.
\newblock doi:{10.1371/journal.pone.0259166}.

\bibitem{gostic_practical_2020}
Gostic KM, McGough L, Baskerville EB, Abbott S, Joshi K, Tedijanto C, et~al.
\newblock Practical considerations for measuring the effective reproductive number, {Rt}.
\newblock PLOS Computational Biology. 2020;16(12):e1008409.
\newblock doi:{10.1371/journal.pcbi.1008409}.

\bibitem{richardson_tracking_2022}
Richardson R, Jorgensen E, Arevalo P, Holden TM, Gostic KM, Pacilli M, et~al.
\newblock Tracking changes in {SARS}-{CoV}-2 transmission with a novel outpatient sentinel surveillance system in {Chicago}, {USA}.
\newblock Nature Communications. 2022; p. 5547.
\newblock doi:{10.1038/s41467-022-33317-6}.

\bibitem{goldstein_semiparametric_2024}
Goldstein IH, Parker DM, Jiang S, Minin VM. Semiparametric inference of effective reproduction number dynamics from wastewater pathogen surveillance data; 2024.
\newblock Available from: \url{http://arxiv.org/abs/2308.15770}.

\bibitem{wilkinson2016fair}
Wilkinson MD, Dumontier M, Aalbersberg IJ, Appleton G, Axton M, Baak A, et~al.
\newblock {The FAIR Guiding Principles for scientific data management and stewardship}.
\newblock Scientific data. 2016;3(1):1--9.

\bibitem{grinberg2018flask}
Grinberg M.
\newblock Flask web development.
\newblock O'Reilly Media, Inc.; 2018.

\bibitem{ananthakrishnan2018globus}
Ananthakrishnan R, Blaiszik B, Chard K, Chard R, McCollam B, Pruyne J, et~al.
\newblock Globus platform services for data publication.
\newblock In: Proceedings of the Practice and Experience on Advanced Research Computing: Seamless Creativity; 2018. p. 1--7.

\bibitem{correct}
Hayot-Sasson V. CORRET; 2025.
\newblock \url{https://github.com/globus-labs/correct}.

\bibitem{ur2014practical}
Ur B, McManus E, Pak Yong~Ho M, Littman ML.
\newblock Practical trigger-action programming in the smart home.
\newblock In: Proceedings of the SIGCHI conference on human factors in computing systems; 2014. p. 803--812.

\bibitem{foster2011globus}
Foster I.
\newblock {Globus Online: Accelerating and democratizing science through cloud-based services}.
\newblock IEEE Internet Computing. 2011;15(3):70--73.

\bibitem{chard20funcx}
Chard R, Babuji Y, Li Z, Skluzacek T, Woodard A, Blaiszik B, et~al.
\newblock func{X}: A Federated Function Serving Fabric for Science.
\newblock In: Proceedings of the 29th International Symposium on High-Performance Parallel and Distributed Computing. ACM; 2020.Available from: \url{http://dx.doi.org/10.1145/3369583.3392683}.

\bibitem{ananthakrishnan24multiuser}
Ananthakrishnan R, Babuji Y, Baughman M, Bryan J, Chard K, Chard R, et~al.
\newblock Enabling Remote Management of {FaaS} Endpoints with {Globus Compute} Multi-User Endpoints.
\newblock In: Practice and Experience in Advanced Research Computing 2024: Human Powered Computing. PEARC '24. New York, NY, USA: Association for Computing Machinery; 2024.Available from: \url{https://doi.org/10.1145/3626203.3670612}.

\bibitem{keahey2020lessons}
Keahey K, Anderson J, Zhen Z, Riteau P, Ruth P, Stanzione D, et~al.
\newblock Lessons Learned from the Chameleon Testbed.
\newblock In: Proceedings of the 2020 USENIX Annual Technical Conference (USENIX ATC '20). USENIX Association; 2020.

\bibitem{huisman_estimation_2022}
Huisman JS, Scire J, Angst DC, Li J, Neher RA, Maathuis MH, et~al.
\newblock Estimation and worldwide monitoring of the effective reproductive number of {SARS}-{CoV}-2.
\newblock eLife. 2022;11.

\bibitem{epyestim}
Hilfiker L, Josi J. epyestim; 2021.
\newblock \url{https://pypi.org/project/epyestim/}.

\bibitem{cori_new_2013}
Cori A, Ferguson NM, Fraser C, Cauchemez S.
\newblock A New Framework and Software to Estimate Time-Varying Reproduction Numbers During Epidemics.
\newblock American Journal of Epidemiology. 2013; p. 1505--1512.
\newblock doi:{10.1093/aje/kwt133}.

\bibitem{goldstein_reconstructing_2009}
Goldstein E, Dushoff J, Ma J, Plotkin JB, Earn DJD, Lipsitch M.
\newblock Reconstructing influenza incidence by deconvolution of daily mortality time series.
\newblock Proceedings of the National Academy of Sciences. 2009;106(51):21825--21829.
\newblock doi:{10.1073/pnas.0902958106}.

\bibitem{goldstein_semiparametric_2023}
Goldstein IH, Parker DM, Jiang S, Minin VM.
\newblock Semiparametric inference of effective reproduction number dynamics from wastewater pathogen surveillance data.
\newblock Biometrics. 2024;80(3).
\newblock doi:{10.1093/biomtc/ujae074}.

\bibitem{vassiliadis2009survey}
Vassiliadis P.
\newblock A survey of extract--transform--load technology.
\newblock International Journal of Data Warehousing and Mining. 2009;5(3):1--27.

\bibitem{Saxena2023}
Saxena M, Sowell B, Alamgir D, Bahadur N, Bisht B, Chandrachood S, et~al.
\newblock {The story of AWS Glue}.
\newblock In: VLDB 2023; 2023.Available from: \url{https://www.amazon.science/publications/the-story-of-aws-glue}.

\bibitem{rawat2019introduction}
Rawat S, Narain A, Rawat S, Narain A.
\newblock {Introduction to Azure Data Factory}.
\newblock Understanding Azure Data Factory: Operationalizing Big Data and Advanced Analytics Solutions. 2019; p. 13--56.

\bibitem{brinckman2019computing}
Brinckman A, Chard K, Gaffney N, Hategan M, Jones MB, Kowalik K, et~al.
\newblock {Computing environments for reproducibility: Capturing the “Whole Tale”}.
\newblock Future Generation Computer Systems. 2019;94:854--867.

\bibitem{appukuttan2023ebrains}
Appukuttan S, Bologna LL, Sch{\"u}rmann F, Migliore M, Davison AP.
\newblock {EBRAINS live papers--interactive resource sheets for computational studies in neuroscience}.
\newblock Neuroinformatics. 2023;21(1):101--113.

\bibitem{ellerm22live}
Ellerm A, Adams B, Gahegan M, Trombach L.
\newblock Enabling {LivePublication}.
\newblock In: 18th International Conference on e-Science (e-Science); 2022. p. 419--420.

\bibitem{sanz2022neuroci}
Sanz-Robinson J, Jahanpour A, Phillips N, Glatard T, Poline JB.
\newblock Neuro{CI}: Continuous Integration of Neuroimaging Results Across Software Pipelines and Datasets.
\newblock In: 2022 IEEE 18th International Conference on e-Science (e-Science). IEEE; 2022. p. 105--116.

\bibitem{poline2023data}
Poline JB, Das S, Glatard T, Madjar C, Dickie EW, Lecours X, et~al.
\newblock {Data and tools integration in the Canadian Open Neuroscience Platform}.
\newblock Scientific Data. 2023;10(1):189.

\bibitem{sage}
{Sage App Catalogue}; 2024.
\newblock \url{https://portal.sagecontinuum.org/apps/explore}.

\bibitem{pegasushub}
{PegasusHub}; 2024.
\newblock \url{https://pegasushub.io}.

\bibitem{huggingface}
{HuggingFace}; 2024.
\newblock \url{https://huggingface.co}.

\bibitem{kaggle}
{Kaggle}; 2024.
\newblock \url{https://www.kaggle.com}.

\bibitem{markiewicz2021openneuro}
Markiewicz CJ, Gorgolewski KJ, Feingold F, Blair R, Halchenko YO, Miller E, et~al.
\newblock {The OpenNeuro resource for sharing of neuroscience data}.
\newblock Elife. 2021;10:e71774.

\bibitem{clarke2017international}
Clarke L, Fairley S, Zheng-Bradley X, Streeter I, Perry E, Lowy E, et~al.
\newblock {The international Genome sample resource (IGSR): A worldwide collection of genome variation incorporating the 1000 Genomes Project data}.
\newblock Nucleic acids research. 2017;45(D1):D854--D859.

\bibitem{cdp}
{Chicago Data Portal}; 2024.
\newblock \url{https://data.cityofchicago.org/}.

\bibitem{wagner2022fairly}
Wagner AS, Waite LK, Wierzba M, Hoffstaedter F, Waite AQ, Poldrack B, et~al.
\newblock {FAIRly big: A framework for computationally reproducible processing of large-scale data}.
\newblock Scientific Data. 2022;9(1):80.

\bibitem{halchenko2021datalad}
Halchenko Y, Meyer K, Poldrack B, Solanky D, Wagner A, Gors J, et~al.
\newblock Data{L}ad: Distributed system for joint management of code, data, and their relationship.
\newblock Journal of Open Source Software. 2021;6(63).

\bibitem{sherif2014cbrain}
Sherif T, Rioux P, Rousseau ME, Kassis N, Beck N, Adalat R, et~al.
\newblock {CBRAIN: a web-based, distributed computing platform for collaborative neuroimaging research}.
\newblock Frontiers in neuroinformatics. 2014;8:54.

\bibitem{stubbs2021tapis}
Stubbs J, Cardone R, Packard M, Jamthe A, Padhy S, Terry S, et~al.
\newblock Tapis: An {API} platform for reproducible, distributed computational research.
\newblock In: Future of Information and Communication Conference, Volume 1. Springer; 2021. p. 878--900.

\bibitem{hedges2007management}
Hedges M, Hasan A, Blanke T.
\newblock {Management and preservation of research data with iRODS}.
\newblock In: Proceedings of the ACM first workshop on CyberInfrastructure: information management in eScience; 2007. p. 17--22.

\bibitem{chard2017ripple}
Chard R, Chard K, Alt J, Parkinson DY, Tuecke S, Foster I.
\newblock Ripple: Home automation for research data management.
\newblock In: 37th International Conference on Distributed Computing Systems Workshops. IEEE; 2017. p. 389--394.

\end{thebibliography}
